\documentclass[11pt]{article}

\usepackage{amssymb}
\usepackage{amsmath}
\usepackage{amsthm}
\usepackage{color}
\usepackage{graphicx}
\usepackage{hyperref}
\usepackage{fullpage}
\usepackage{graphicx}
\usepackage{mathtools}
\usepackage{microtype}
\usepackage{bbm}
\usepackage{multirow}
 \usepackage[normalem]{ulem}
\useunder{\uline}{\ul}{}
\usepackage[linesnumbered,algoruled,titlenotnumbered,vlined]{algorithm2e}
\usepackage{indentfirst}
\DontPrintSemicolon

\usepackage{algorithmic}

\usepackage{lmodern}
\usepackage[T1]{fontenc}

\newtheorem{theorem}{Theorem}
\newtheorem{lemma}[theorem]{Lemma}
\newtheorem{definition}[theorem]{Definition}

\newtheorem{corollary}[theorem]{Corollary}

\newtheorem{proposition}[theorem]{Proposition}

\renewenvironment{proof}{\vspace{-0.05in}\noindent{\bf Proof:}}%
        {\hspace*{\fill}$\Box$\par}
\newenvironment{proofof}[1]{\smallskip\noindent{\bf Proof of #1:}}%
        {\hspace*{\fill}$\Box$\par}
        {\hspace*{\fill}$\Box$\par}

\newcommand{\ie}{{\em i.e.}}

\DeclareMathOperator{\poly}{poly}

\DeclareMathOperator{\argmax}{argmax}
\DeclareMathOperator{\argmin}{argmin}

\def\({\left(}
\def\){\right)}

\begin{document}
\title{Nearly Linear Time Deterministic Algorithms for Submodular Maximization Under Knapsack Constraint and Beyond}

\author{Wenxin Li \\
 The Ohio State University\\
{\tt wenxinliwx.1@gmail.com}\\
}
\maketitle

\begin{abstract}
In this work, we study the classic submodular maximization problem under knapsack constraints and beyond. We first present an $(7/16-\varepsilon)$-approximate algorithm for single knapsack constraint, which requires $O(n\cdot\max\{\varepsilon^{-1},\log\log n\})$ queries, and two passes in the streaming setting. This provides improvements in approximation ratio, query complexity and number of passes on the stream. We next show that there exists an $(1/2-\varepsilon)$-approximate deterministic algorithm for constant number of binary packing constraints, which achieves a query complexity of $O_{\varepsilon}(n\cdot\log \log n)$. One salient feature of our deterministic algorithm is, both its approximation ratio and time complexity are independent of the number of constraints. Lastly we present nearly linear time algorithms for the intersection of $p$-system and $d$ knapsack constraint, we achieve approximation ratio of $(1/(p+\frac{7}{4}d+1)-\varepsilon)$ for monotone objective and $(\frac{p/(p+1)}{2p+\frac{7}{4}d+1}-\varepsilon)$ for non-monotone objective.
\end{abstract}


\section{Introduction}
A set function $f:2^{E}\rightarrow \mathbb{R}^{+}$ defined on ground $E$ of size $n$ is \emph{submodular}, if for any two sets $S,T\subseteq E$, inequality $f(S)+f(T)\geq f(S\cup T)+f(S\cap T)$ holds. It is \emph{monotone non-decreasing} if $f(S)\leq f(T)$ for any $S\subseteq T\subseteq E$. Submodular functions form a natural class of set functions with the property of diminishing returns, which have numerous applications in computer science, economics, and operation research. Due to its widespread applicability of submodular maximization, there has been a vast amount of literature on submodular maximization subject to diverse types of constraints~\cite{nemhauser1978analysis, khuller1999budgeted, sviridenko2004note,calinescu2011maximizing,feldman2011unified,DBLP:journals/siamdm/Yoshida19,filmus2012tight,chekuri2015multiplicative}.

However, many of these algorithmic results do not scale well for practical applications of large size. Obtaining fast running time is of fundamental importance in both theory and practice [23] and there has been a considerable amount of work in this direction. Traditionally, a linear query complexity algorithm for a problem is highly desirable. Hence one question arising is, what is the best solution that can be obtained in (nearly) linear time? In this paper we aim to understand the approximation boundary via (nearly) linear number of queries. In the mean time, we also try to reduce the previous query complexity. For example, in several of our results $O_{\varepsilon}(n\log\log n)$ queries are required, while $\Omega_{\varepsilon}(\log n)$ queries per element are required in previous works, which is exponentially higher than the complexity result(s) in this paper.

A large number of applications are formulated as knapsack constrained monotone submodular function maximization problem. Sviridenko~\cite{sviridenko2004note} proposed the density greedy algorithm with partial enumeration to obtain the optimal approximation ratio of $1-1/e$. The algorithm requires time complexity of $O(n^{5})$ and is computationally inefficient. In recent years there has been a large amount of work focusing on solving the knapsack constrained submodular maximization problem in large-scale scenarios. For example, for the line of algorithm acceleration initialized by~\cite{badanidiyuru2014fast}, the current best result is due to Ene and Nguyen~\cite{ene2017nearly}, in which a randomized $O((1/\varepsilon)^{O(1/\varepsilon^{4})}\cdot n\log^{2} n)$ time algorithm with approximation factor $(1-1/e-\varepsilon)$ was proposed. When working with massive data stream, Huang et al.~\cite{huang2019streaming} proposed an $(0.363-\varepsilon)$-approximate single pass streaming algorithm, which requires $O((1/\varepsilon^{4})\log^{4}n)$ space and queries per element, together with a $(0.4-\varepsilon)$-approximate three pass algorithm with the same space and running time requirements. Motivated by \emph{reducing the number of passes on the data stream}, together with the \emph{gap between the unnatural state-of-the-art and existing hardness result}, we present our improved algorithm in Theorem~\ref{budgettheorem}.

We further investigate other types of knapsack constraints, binary packing constraint. For multiple packing constraints, Azar and Gamzu~\cite{azar2012efficient} proposed a multiplicative weight update (MWU)-based greedy algorithm that achieves a width-dependent approximation ratio of $\Omega(1/d^{1/W})$, where $d$ represents the number of constraints and $W$ refers to the width of the packing system. 
The approximation guarantee can be further improved to $\Omega(1/d^{1/W+1})$ for binary packing constraints. While the aforementioned approximation guarantees are dependent on the number of constraints, Mizrachi et al.~\cite{mizrachi2019tight} proposed the first deterministic non-trivial algorithm for a constant number of packing constraints. However, it requires $O(n^{O(\poly (1/\varepsilon))})$ time to achieve an approximation ratio of $(1/e-\varepsilon)$. One question we ask is, \emph{does there exist a time efficient deterministic algorithm with width independent approximation ratio?} We answer this in affirmative for constant number of binary packing constraints in Theorem~\ref{binarysingleknap}.

The most general type of constraint considered in this paper is the intersection of a $p$-system and $d$ knapsack constraints \cite{badanidiyuru2014fast}. The current best result is due to Badanidiyuru and Vondr{\'a}k~\cite{badanidiyuru2014fast}, in which an $(1/(p+2d+1)-\varepsilon)$-approximate algorithm was proposed. On the inapproximability side, there is a lower bound of $(1-e^{-p})+\varepsilon \leq 1/(p+1/2)+\varepsilon$ even for the special case of $p$--extendible system~\cite{feldman2017greed}. To the best of our knowledge, there is no algorithm that is able to move closer towards the lower bound by using comparable number of queries. We provide a nearly linear algorithm with better approximation guarantee in Theorem~\ref{psystemtheorem}.

\subsection{Results Overview}\label{mainresult}

\paragraph{Knapsack constraint.} In Section~\ref{knapsacksec} we study single knapsack constrained submodular maximization problem and have the following theorem.


\begin{theorem}\label{budgettheorem}
There is an $(7/16-\varepsilon)$-approximate algorithm for maximizing a monotone submodular function subject to a single knapsack constraint, which requires $O(\max\{\varepsilon^{-1},\log\log n\})$ queries per element. Our algorithm can be adapted to the streaming setting, in which the same approximation ratio can be achieved within only \textbf{two} passes over the data stream, while using $O(n\log n/\varepsilon)$ space and performing $O(\log n/\varepsilon)$ queries per element. 
\end{theorem}

We improve the $(0.4-\varepsilon)$-approximate algorithm in~\cite{huang2017streaming,huang2019streaming}, which requires $O(\log^{4}n/\varepsilon^{4})$ queries per element, $O(n\log^{4}n/\varepsilon^{4})$ space and three passes on the stream. We note that the algorithms in~\cite{huang2018m} require a larger number of passes on the data stream to achieve the same approximation ratio.  Our next result applies to binary packing constraint with constant dimension.

We also study the case when the constraint set $\mathcal{I}=\{S\subseteq E|\mathbf{A}\mathbf{x}_{S}\leq \mathbf{b}\}$. More specifically, we investigate two specific forms of packing constrained optimization problems. The first one is the well-known single knapsack constraint. 
\begin{theorem}\label{binarysingleknap}
There is an $(1/2-\varepsilon)$-approximate deterministic algorithm
that performs $O_{\varepsilon}(n\cdot\log\log n)$ queries for the binary packing constraint with constant dimension, \ie, $\mathbf{A}\in \{0,1\}^{d\times n}$ where $d= O(1)$. The result also holds if $\mathbf{A}\in \mathbb{F}^{d\times n}$, where $\mathbb{F}$ consists of constants and $|\mathbb{F}|=O(1)$.
\end{theorem} 
We would like to emphasize that our approximation ratio is width independent and holds deterministically. Compared with the algorithm in~\cite{mizrachi2019tight}, the complexity of our algorithm for constant number of binary packing constraints is nearly linear in the input size.

\paragraph{Intersection of $p$-system and $d$-knapsack constraints.} In Section~\ref{psystemdknapsection}, we study the problem when the constraint set $\mathcal{I}$ represents the intersection of $p$-system and $d$ knapsack constraints. 

\begin{theorem}\label{psystemtheorem}
There is an $(1/(p+\frac{7}{4}d+1)-\varepsilon)$-approximate algorithm for maximizing a non-negative monotone submodular function subject to a $p$-system and $d$ knapsack constraints, which performs nearly linear number of queries. 
\end{theorem}
We improve the approximation ratio of $(1/(p+2d+1)-\varepsilon)$ in~\cite{badanidiyuru2014fast} and the improvement is constant for constant value of $p$ and $d$. On the other hand, similar as Theorem~\ref{psystemtheorem}, we improve the approximation ratio of $(\frac{p/(p+1)}{2p+2d+1}-\varepsilon)$ in~\cite{mirzasoleiman2016fast} for non-monotone submodular maximization.

\begin{theorem}\label{nonmonotoneknappsys}
There is an $(\frac{p/(p+1)}{2p+\frac{7}{4}d+1}-\varepsilon)$-approximate algorithm for maximizing a (non-monotone) submodular function subject to the intersection of $p$-system and $d$ knapsack constraints, which performs nearly linear number of queries.
\end{theorem}

\section{An Efficient Algorithm for Single Knapsack Constraint}\label{knapsacksec}

Our improved solution for a single knapsack constraint consists of Algorithm~\ref{streamingmain}, a 
backtracking algorithm utilizing multiple thresholds, and Algorithm~\ref{guesslargest1}, an alternative algorithm for the case when there exists an element with cost no less than $1/2$ in $\mathrm{OPT}$. In the following we give an overview of these two procedures. 

\paragraph{Overview of the two subroutines.} We run two threads in parallel in Algorithm~\ref{streamingmain}, the \emph{double threshold backtracking algorithm}, where each thread contains two sequential stages. In stage $j$ of thread $i$ ($i,j\in [2]$), we select elements whose profit density is no less than the profit density threshold $\tau^{(i)}_{j}\in [\varepsilon f(\mathrm{OPT}), c^{(i)}_{j}\cdot f(\mathrm{OPT})]$, and cost no more than $1/i$ (this constraint is trivial in the first thread). We recursively construct a new candidate solution $\widetilde{T}_{j}^{(i)}$ if there exists constraint violation in the proceeding procedure. We output the best solution based on the collection of sets obtained in the aforementioned two threads. In Algorithm~\ref{guesslargest1}, we first select a singleton with both function value and cost close to that of the element with largest cost in $\mathrm{OPT}$, then Algorithm~\ref{mainalgorithmpsystemdknapsack} is utilized to solve the corresponding residual problem. The final solution is chosen to be the best set among solutions to the $O(1/\delta)$ problem and that returned by Algorithm~\ref{streamingmain}. It is important to note that $\delta=\Theta(1)$, according to inequality (\ref{deterdelta}).

\begin{algorithm}[H]\label{guesslargest1}
\small
\caption{Main Algorithm for Single Knapsack Constraint}
\textbf{Initialization:} $\bar{\zeta} \leftarrow f(\mathrm{OPT})$, $\underline{\zeta}\leftarrow \frac{13\bar{\zeta}}{48}$\\
\While {$\zeta \geq \underline{\zeta}$}
{$e_{\zeta}\leftarrow \argmin\{c(e)|\zeta \leq f(e)\leq \frac{\zeta}{1-\delta} ,e\in E\}$\\
\tcp{$\delta$ is a constant independent of the approximation parameter $\varepsilon$}
With budget $1-c(e_{\zeta})$, apply the backtracking threshold algorithm (Algorithm~\ref{mainalgorithmpsystemdknapsack}) on function $g_{\zeta}(\cdot):2^{E\setminus\{e_{\zeta}\}}\rightarrow \mathbb{R}^{+}$ to obtain solution $S_{\zeta}$, where $g_{\zeta}(S)=f(S+e_{\zeta})-f(e_{\zeta})$ for $\forall S\subseteq E\setminus\{e_{\zeta}\}$\\
$\zeta\leftarrow (1-\delta)\cdot \zeta$\\ 
}
$\zeta^{*}\leftarrow \argmax_{\zeta}{f(S_{\zeta}+e_{\zeta})}$\\
$S^{\prime}_{o}\leftarrow S_{\zeta^{*}}+e_{\zeta^{*}}$\\
$S^{*}\leftarrow$ solution returned by Algorithm~\ref{streamingmain}\\
\textbf{Return} $S_{o}\leftarrow \argmax_{S\in \{S^{*},  S^{\prime}_{o}\}}f(S)$
\end{algorithm}

\subsection{An $O(n\cdot\log\log n)$ time $\frac{1}{3e}$-approximate algorithm}

In the $i$-th iteration of the \emph{adaptive decreasing threshold} (ADT) algorithm, we maintain $\bar{w}_{i}$ and $\underline{w}_{i}$ as an upper and lower estimate on the optimal objective value $f(\mathrm{OPT})$. At the end iteration $i$, the lower estimate of $f(\mathrm{OPT})$ is updated as the maximum function value of the sets obtained in $i$-th iteration. It turns out that the gap between the upper and lower estimates of $f(\mathrm{OPT})$ is a constant after $\ell=O(\log \log n)$ iterations. In the following lemma we prove that $\underline{w}_{i}$ and $\bar{w}_{i}$ are always valid lower and upper bounds on $f(\mathrm{OPT})$. We compare ADT with existing algorithm in~\cite{huang2018multi} in Appendix~\ref{adtcompare}.
\begin{algorithm}[H]\label{sizealgoadt}
\small
    \caption{Adaptive Decreasing Threshold (ADT) Algorithm}
\textbf{Initialization:} $\underline{\omega}_1\leftarrow \max_{e\in E}{f(e)}$, $\bar{\omega}_1\leftarrow k\cdot \underline{\omega}_1$, $U\leftarrow \varnothing$, 
$\ell \leftarrow \lceil\log \log n\rceil $.
\\
\For{$i=1:\ell$}
{$\alpha_i=\exp(\log n\cdot e^{-i})-1$, $\theta=\underline{\omega}_{i}$\\
\While{$\theta\leq \bar{\omega}_{i}$}
{$S^{(i)}_{\theta}\leftarrow \varnothing$\\
\For{$e\in E$}
{\If{$c(S^{(i)}_{\theta}+e)>1$}
{$S^{(i)}_{\theta}\leftarrow \argmax_{T\in\{\{S^{(i)}_{\theta}\},\{e\}\}}{f(T)}$
}
\ElseIf{$f(S^{(i)}_{\theta}+e)-f(S^{(i)}_{\theta})\geq \theta\cdot c(e)$}
{$S^{(i)}_{\theta}\leftarrow S^{(i)}_{\theta}+e$
}
}
$\theta\leftarrow \theta(1+\alpha_{i})$
}
$\underline{\omega}_{i+1} \leftarrow \max_{\theta}{f(S^{(i)}_{\theta})}$, $\bar{\omega}_{i+1}\leftarrow 3(1+\alpha_{i})\cdot\underline{\omega}_{i+1}$\\
} 
\Return {$\bar{w}_{\ell}$}
\end{algorithm}

\begin{lemma} \label{sizelemma} For any $i\in [\ell]$, the optimal objective value always lies between $\underline{w}_{i}$ and $\bar{w}_{i}$, \ie, $\underline{w}_{i} \leq f(\mathrm{OPT}) \leq \bar{w}_{i}$. As a consequence, $\frac{f(\mathrm{OPT})}{3e}\leq \underline{w}_{\ell} \leq f(\mathrm{OPT})$.
\end{lemma}
\begin{proof}
We finish the proof by induction. For the base case when $i=1$, as $\underline{w}_{1}$ is initialized to be the maximum objective value of a singleton, Lemma~\ref{sizelemma} is equivalent to $\max_{e}{f(e)}\leq f(\mathrm{OPT}) \leq n\cdot \max_{e}{f(e)}$, which follows from the submodularity of $f(\cdot)$. Notice that for $i\geq 2$, we have $\underline{w}_{i}=\max_{\theta}{f(S^{(i-1)}_{\theta})}$, where $S_{\theta}^{(i-1)}$ is a feasible solution. Hence $\underline{w}_{i}$ is always a valid lower bound of $f(\mathrm{OPT})$ and what remains to prove is $\bar{w}_{i}\geq f(\mathrm{OPT})$ for $\forall i\in [\ell]$.

\paragraph{Induction Step.} Assume that $\bar{w}_{i}\geq f(\mathrm{OPT})$ holds for $i=q$. In the following, we complete the proof for $i=q+1$ by lower bounding the objective value of $f(S^{(q)}_{\theta^{*}_{q}})$. Observe that in the $q$-th iteration, $\theta$ takes values in set 
\begin{align*}
\Theta_{q}=\Big\{\underline{w}_{q}, \underline{w}_{q}(1+\alpha_{q}),\ldots, \underline{w}_{q}(1+\alpha_{q})^{\lfloor \log (\bar{w}_{q}/ \underline{w}_{q})/{\log (1+\alpha_{q})} \rfloor}\Big\}.
\end{align*}
Combined with the induction assumption $\bar{w}_{q}\geq f(\mathrm{OPT})$, there must exist some $\theta^{*}_{q}\in \Theta_{q}$ such that $\theta^{*}_{q}\leq \frac{2}{3}f(\mathrm{OPT})\leq (1+\alpha_{q})\theta^{*}_{q}$. Consider the iteration in which $\theta=\theta^{*}_{q}$, let $\mathcal{E}$ denote the event that there exists element $e$ such that $c(S_{\theta^{*}_{q}}^{(q)}+e)>1$, then we can lower bound $f(S_{\theta^{*}_{q}}^{(q)})$ by its size multiplying the corresponding threshold, \ie, 
\begin{align*}
\max\Big\{f(S^{(q)}_{\theta^{*}_{q}}),f(e)\Big\}\geq \frac{f(S^{(q)}_{\theta^{*}_{q}}+e)}{2}\mathbbm{1}_{\bar{\mathcal{E}}}\geq  \theta^{*}_{q}\cdot \frac{c(S^{(q)}_{\theta^{*}}+e)}{2}\cdot \mathbbm{1}_{\bar{\mathcal{E}}}\geq \frac{\theta^{*}_{q}}{2}\cdot \mathbbm{1}_{\bar{\mathcal{E}}}\geq \frac{f(\mathrm{OPT})}{3(1+\alpha_{q})}\cdot \mathbbm{1}_{\bar{\mathcal{E}}}.
\end{align*}
If there exists no element exceeding the budget, then elements in $\mathrm{OPT}\setminus S^{(q)}_{\theta^{*}_{q}}$ will have a small marginal gain with respect to $S^{(q)}_{\theta^{*}_{q}}$ and
\begin{align*}
&[f(\mathrm{OPT})-f(S^{(q)}_{\theta^{*}_{q}})]\cdot \mathbbm{1}_{\bar{\mathcal{E}}} \leq f(\mathrm{OPT}\cup S^{(q)}_{\theta^{*}_{q}})-f(S^{(q)}_{\theta^{*}_{q}})\tag{monotonicity}\\
\leq &\sum_{e\in \mathrm{OPT}}{\Big[f(S^{(q)}_{\theta^{*}_{q}}+e)-f(S^{(q)}_{\theta^{*}_{q}})\Big]}\leq \sum_{e\in \mathrm{OPT}}{\theta^{*}_{q}\cdot c(e)}=\theta^{*}_{q}\leq \frac{2f(\mathrm{OPT})}{3}. 
\end{align*}
This implies that $f(S^{(q)}_{\theta^{*}})\geq \frac{f(\mathrm{OPT})}{3}\cdot \mathbbm{1}_{\bar{\mathcal{E}}}$. To summarize, we have 
\begin{align}\label{lastineq}
f(S^{(q)}_{\theta^{*}})\geq \frac{f(\mathrm{OPT})}{3(1+\alpha_{q})}\cdot \mathbbm{1}_{\mathcal{E}}+\frac{f(\mathrm{OPT})}{3}	\cdot \mathbbm{1}_{\bar{\mathcal{E}}} \geq \frac{f(\mathrm{OPT})}{3(1+\alpha_{q})}.
\end{align}
Therefore $\bar{w}_{q+1}=3(1+\alpha_{q})\underline{w}_{q+1}=3(1+\alpha_{q})
\max_{\theta}{f(S^{(q)}_{\theta})}\geq 3(1+\alpha_{q})f(S^{(q)}_{\theta^{*}_{q}})\geq f(\mathrm{OPT})$, which are mainly based on (\ref{lastineq}) and the definition of $\underline{w}_{i}$, $\bar{w}_{i}$. For $i=\ell$ we have 
\begin{align*}
\bar{w}_{\ell}=(1+\alpha_{\ell})\underline{w}_{\ell}=\exp (\log n\cdot e^{-\lceil\log \log n \rceil})\underline{w}_{\ell}\leq e\underline{w}_{\ell}.
\end{align*}
The proof is complete.
\end{proof}


\begin{proposition}[Complexity of Algorithm~\ref{sizealgoadt}]\label{ADTtime}
Algorithm~\ref{sizealgoadt} performs $O(n\log\log n)$ queries in total.
\end{proposition}
\begin{proof} Note that in the $i$-th iteration of the preprocessing procedure, we perform $O\Big(n\cdot\frac{\log {(\underline{\omega}_{i}/\bar{\omega}_{i}) }}{\log (1+\alpha_{i})}\Big)= O\Big(n\cdot\frac{\log{(1+\alpha_{i-1}})}{\log(1+\alpha_{i})}\Big)$ number of queries, which implies that the total number of queries performed is in the order of 
\begin{align*}
O\Big(\sum_{i=1}^{\ell}{\frac{\log {(1+\alpha_{i-1}})}{\log(1+\alpha_{i})}}\Big)=O\Big(\sum_{i=1}^{\ell}{\frac{\log(\exp(\log k\cdot e^{-i+1}))}{\log(\exp(\log k\cdot e^{-i}))}}\Big)=O(e\ell)= O(\log\log n).
\end{align*}
\end{proof}

\subsection{Double threshold backtracking algorithm and performance analysis~\ref{streamingmain}}

\begin{algorithm}[H]\label{streamingmain}
\small
    \caption{Double Threshold Backtracking Algorithm}
\textbf{Initialization:} $q\leftarrow 2$,  $T^{(i)}_{0}, T^{(i)}_{j}, \widetilde{T}_{j}^{(i)}\leftarrow \varnothing \;(\forall i, j\in [q])$, $\lambda \leftarrow f(\mathrm{OPT})$\\ 
\While{$\lambda\geq \varepsilon \cdot f(\mathrm{OPT})$}
{\tcp{A constant approximation of $f(\mathrm{OPT})$ is sufficient for initializing $\lambda$}
\For{$ i \in  [q]$}
{
{\For{$j\in [q]$}
{$T^{(i)}_{j}\leftarrow T^{(i)}_{j-1}$, $c^{(i)}_{1}\leftarrow \frac{3i}{3+i}$, $c^{(i)}_{2}\leftarrow \frac{9}{(3+i)^{2}}$, $\tau^{(i)}_{j}\leftarrow \lambda\cdot c^{(i)}_{j}(j\in [q])$\\ 
\For{each $e\in E\setminus T^{(i)}_{j}$}
{\If{$f(T^{(i)}_{j}+e)-f(T^{(i)}_{j})\geq c(e)\cdot \tau^{(i)}_{j}$ and $c(e)\leq 1/i$}
{
\If{$c(T^{(i)}_{j}+e)\leq 1$}
{$T^{(i)}_{j}\leftarrow T^{(i)}_{j}+e$\\}
\Else
{ $\hat{e}^{(i)}_{j}\leftarrow e$, $\widetilde{T}_{j}^{(i)} \leftarrow \widetilde{T}_{j}^{(i)}+e$\\
\For{$e\in T^{(i)}_{j}$}
{\If{$c(\widetilde{T}_{j}^{(i)}+e)\leq 1$}
{$\widetilde{T}_{j}^{(i)} \leftarrow \widetilde{T}_{j}^{(i)}+e$\\}
}
}
}

}
}
}
\For{$i\in [q]$}
{
$\tilde{e}^{(i)}_{1}\leftarrow \argmax_{e\in T^{(i)}}{c(e)}$, $\tilde{e}^{(i)}_{2}\leftarrow \argmax_{e\in T^{(i)}_{2}\setminus T^{(i)}_{1}}{c(e)}$\\
$T^{(i)}\leftarrow \cup_{j\in [q]}{T^{(i)}_{j}}$, $U^{(i)}_{1}\leftarrow \{\hat{e}^{(i)},\tilde{e}^{(i)}_{1}\}$, $U^{(i)}_{2}\leftarrow\{\hat{e}^{(i)}, \tilde{e}^{(i)}_{2}\}$, $U^{(i)}_{3}\leftarrow\{\hat{e}^{(i)}, \tilde{e}^{(i)}_{2}\}\cup T^{(i)}_{1}$
}
}
$\lambda \leftarrow (1-\varepsilon)\cdot \lambda$\\
}
\textbf{Return} $S^{*}\leftarrow \argmax{\{f(S)|S\in \{U^{(i)}_{\ell}\}_{1\leq \ell\leq 3}\cup \{T^{(i)},\widetilde{T}^{(i)}\}_{i\in [q]}, c(S)\leq 1\}}$  
\end{algorithm}

\begin{lemma}\label{propositionmpass}
For set $S^{*}$ returned by Algorithm~\ref{streamingmain}, its objective value satisfies
\begin{align*}
f(S^{*})\geq \frac{7}{16} f(\mathrm{OPT}) \cdot \mathbbm{1}_{\mathrm{OPT}\cap B=\varnothing}+\frac{16}{25}[f(\mathrm{OPT})-f(\mathrm{OPT}\cap B)]\cdot \mathbbm{1}_{\mathrm{OPT}\cap B\neq \varnothing}-O(\varepsilon \cdot f(\mathrm{OPT})),
\end{align*}	
where $B=\{e\in E|c(e)\geq \frac{1}{2}\}$ represents the set of large elements, \ie, elements with cost no less than $1/2$.
\end{lemma}

\begin{proof}
In thread $i$ of Algorithm~\ref{streamingmain}, two thresholds $\tau^{(i)}_{1}$ and $\tau^{(i)}_{2}$ are utilized to select elements. For a clean presentation, we omit the index of the thread and lower bound the quality of solution obtained by two sequential threshold $\tau_{1}$ and $\tau_{2}$ in double threshold backtracking algorithm.

In the following we use $T_{i} \;(1\leq i\leq 2)$ to denote the collection of elements obtained by threshold $\tau_{i}$. If there exists element $\hat{e}_{i}$ that has a marginal increment no less than $\tau_{i}$ but violates the knapsack constraint, $T_{i}$ represents the value of candidate set before element $\hat{e}_{i}$ arrives. We further let  $\mathrm{OPT}^{\prime}=\mathrm{OPT}\setminus B$ and $\tilde{e}_{1}$, $\tilde{e}_{2}$ be the element with largest cost in $T_{1}$ and $T_{2}\setminus T_{1}$ respectively, \ie, 
$\tilde{e}_{1}=\argmax_{e\in T_{1}}{c(e)}$ and $\tilde{e}_{2}=\argmax_{e\in T_{2}\setminus T_{1}}{c(e)}$.

We divide our analysis into three cases, according to the existence of budget violation in each iteration.

\subparagraph*{Case $1$: Algorithm~\ref{streamingmain} stops at $\tau_{1}$.} In this case, there exists $\hat{e}_{1}$ such that $c(T_{1}+\hat{e}_{1})>1$, and the marginal increment of $\hat{e}_{1}$ is no less than $f(T_{1}+\hat{e}_{1})-f(T_{1})\geq \tau_{1}$. Hence the objective value of $\{\hat{e}_{1},\tilde{e}_{1}\}$ can be lower bounded as,
\begin{align*}
f(\{\hat{e}_{1},\tilde{e}_{1}\})\geq f(\hat{e}_{1})+[f(T_{1}+\hat{e}_{1})-f(T_{1})]\geq \tau_{1}(c(\hat{e}_{1})+c(\tilde{e}_{1})).
\end{align*}
According to the definition of $\tilde{e}_{1}$, the cost of set $\widetilde{T}_{1}$ is no less than $1-c(\tilde{e}_{1})$, we have
\begin{align*}
f(\widetilde{T}_{1})\geq \tau_{1}(1-c(\tilde{e}_{1})).
\end{align*}
Combining with the fact that $f(T_{1})\geq \tau_{1}\cdot c(T_{1})\geq \tau_{1}(1-c(\hat{e}_{1}))$. Therefore
\begin{align}\label{tau1vio}
f(S^{*})\geq &\max\{f(T_{1}),f(\{\hat{e}_{1},\tilde{e}_{1}\}), f(\widetilde{T}_{1})\}\notag\\
\geq& \frac{f(T_{1})+f(\hat{e}_{1},\tilde{e}_{1})+f(\widetilde{T}_{1})}{3}\geq \frac{2}{3}\tau_{1}\geq \frac{\tau_{1}+\tau_{2}}{3}.	
\end{align}

\subparagraph*{Case $2$: Algorithm~\ref{streamingmain} stops at $\tau_{2}$.} Without loss of generality, we can assume that $c(T_{1})\leq \frac{2}{3}$, otherwise we can immediately obtain the same lower bound as (\ref{tau1vio}), \ie, 
\begin{align*}
f(S^{*})\geq f(T_{1})\geq c(T_{1})\cdot\tau_{1}\geq \frac{\tau_{1}+\tau_{2}}{3}.    
\end{align*}
With the condition that $c(T_{1})\leq \frac{2}{3}$, the weight of $\hat{e}_{2}$ satisfies that $c(\hat{e}_{2})\geq (1-c(T_{1}))\cdot \mathbbm{1}_{c(T_{1}+\hat{e}_{2})>1}\geq \frac{1}{3}\cdot \mathbbm{1}_{c(T_{1}+\hat{e}_{2})>1}$. Recall that $\widetilde{T}_{2}$ is obtained by adding $\tilde{e}_{2}$ and then elements in $T_{2}$, until the total weight exceeds the budget, we have
\begin{align}
c(\widetilde{T}_{2}-\hat{e}_{2})\geq [1-c(\hat{e}_{2})-c(\tilde{e}_{1})] \cdot \mathbbm{1}_{c(T_{1}+\hat{e}_{2})>1} + [1-c(\hat{e}_{2})-c(\tilde{e}_{2})]\cdot \mathbbm{1}_{c(T_{1}+\hat{e}_{2})\leq 1},
\end{align}
based on which we can obtain the following lower bound on the objective value of $\widetilde{T}_{2}$,
\begin{align}\label{case2ineqtilde}
f(\widetilde{T}_{2})=&[f(\widetilde{T}_{2})-f(\widetilde{T}_{2}-\hat{e}_{2})]+f(\widetilde{T}_{2}-\hat{e}_{2})\notag\\
\geq & [\tau_{2}\cdot c(\hat{e}_{2})+\tau_{1}\cdot c(\widetilde{T}_{2}-\hat{e}_{2})]\cdot \mathbbm{1}_{c(T_{1}+\hat{e}_{2})>1}\notag\\
\geq & [\tau_{2}\cdot c(\hat{e}_{2})+\tau_{1}\cdot (1-c(\hat{e}_{2})-c(\tilde{e}_{1}))	]\cdot \mathbbm{1}_{c(T_{1}+\hat{e}_{2})>1}.
\end{align}
Notice that
\begin{align}\label{case2ineq2ele}
f(\{\hat{e}_{2},\tilde{e}_{1}\})=&[f(\{\hat{e}_{2},\tilde{e}_{1}\})-f(e_{1})]+f(e_{1})\notag\\
\geq &[f(\widetilde{T}_{2}+\hat{e}_{2})-f(\widetilde{T}_{2})]+f(e_{1})\notag\\
\geq& \tau_{2}\cdot c(\hat{e}_{2})+\tau_{1}\cdot c(\tilde{e}_{1}).
\end{align}
Combining (\ref{case2ineqtilde}) and (\ref{case2ineq2ele}) together, we have
\begin{align}
f(S^{*})\geq &\frac{f(\widetilde{T}_{2})+f(\{\hat{e}_{2},\tilde{e}_{1}\})}{2}\notag\\
\geq & \Big[\tau_{2}\cdot c(\hat{e}_{2})+\frac{\tau_{1}}{2}\cdot[1-c(\hat{e}_{2})]\Big]\cdot \mathbbm{1}_{c(T_{1}+\hat{e}_{2})>1}\label{combine222}\\
\geq & \frac{\tau_{1}+\tau_{2}}{3}\cdot \mathbbm{1}_{c(T_{1}+\hat{e}_{2})>1, 2\tau_{2}\geq \tau_{1}},\label{combine2}	
\end{align}
where the last inequality holds due to the the monotonicity of (\ref{combine222}) with respect to $c(\hat{e}_{2})$, together with the fact that $c(\hat{e}_{2})\geq \frac{1}{3}$.

Now we consider the case when $c(T_{1}+\hat{e}_{2})\leq 1$. Due to the definition of $\hat{e}_{2}$, we have $c(T_{2}+\hat{e}_{2})=c(T_{1})+c(T_{2}\setminus T_{1})+c(\hat{e}_{2})>1$, which implies that  $\max\{c(T_{2}\setminus T_{1}),c(\hat{e}_{2})\}\geq \frac{1-c(T_{1})}{2}$. In addition,
\begin{align}\label{3thresholdineqcase22}
f(S^{*})\geq &\max\{f(T_{2}),f(\widetilde{T}_{2})\}\geq 	\max\{f(T_{2}),f({T}_{1}+\hat{e}_{2})\}\notag\\
\geq & f(T_{1})+\max\{f(T_{2})-f(T_{1}),f({T}_{1}+\hat{e}_{2})-f(T_{1})\}\notag\\
\geq & f(T_{1})+\tau_{2}\cdot\max\{c(T_{2}\setminus T_{1}),c(\hat{e}_{2})\}\notag\\
\geq & \tau_{1}\cdot c(T_{1})+\tau_{2}\cdot \frac{1-c(T_{1})}{2}.
\end{align}
Notice that the lower bound in RHS of (\ref{3thresholdineqcase22}) is monotonically increasing with respect to the total weights of $T_{1}$, we have 
\begin{align*}
f(S^{*})\geq \frac{\tau_{1}+\tau_{2}}{3}\cdot \mathbbm{1}_{c(T_{1}+\hat{e}_{2})\leq 1, c(T_{1})\geq \frac{1}{3}}.   
\end{align*}
For the case when $c(T_{1})\leq \frac{1}{3}$, we first argue that $T_{2}\neq T_{1}$. Because the total weights of elements selected in the second iteration is no less than $c(T_{2}\setminus T_{1})> 1-c(T_{1})-c(\hat{e}_{2})\geq \frac{1}{6}>0$, hence element $\tilde{e}_{2}$ must exist. We next claim the following lower bound on $f(S^{*})$,
\begin{align}
f(S^{*})&\geq \max\{f(T_{2}),f(\widetilde{T}_{2})\}\notag\\
&\geq \max\{f(T_{1}+\tilde{e}_{2}),f(T_{1}+\hat{e}_{2})\}\tag{monotonicity of $f(\cdot)$ and $T_{1}+\hat{e}_{2}$ is feasible}\\
&= f(T_{1})+\max\{f(T_{1}+\tilde{e}_{2})-f(T_{1}),f(T_{1}+\hat{e}_{2})-f(T_{1})\}\notag\\
&\geq f(T_{1})+\tau_{2}\cdot\max\{c(\tilde{e}_{2}), c(\hat{e}_{2})\}\label{3threineq}.
\end{align}
Plugging the fact $f(T_{1})\geq f(\mathrm{OPT}^{\prime})-c(\mathrm{OPT}^{\prime})\cdot \tau_{1}$ into (\ref{3threineq}), we have
\begin{align}
f(S^{*})\geq \Big(f(\mathrm{OPT}^{\prime})-c(\mathrm{OPT}^{\prime})\cdot \tau_{1}+\frac{\tau_{2}}{3}\Big)\cdot \mathbbm{1}_{\max\{c(\tilde{e}_{2}), c(\hat{e}_{2})\}\geq \frac{1}{3}}.	
\end{align}
If $\max\{c(\tilde{e}_{2}), c(\hat{e}_{2})\}\leq \frac{1}{3}$, we have $c(T_{1}\cup\{\hat{e}_{2},\tilde{e}_{2}\})\leq 1$, \ie, $T_{1}\cup\{\hat{e}_{2},\tilde{e}_{2}\}$ is a feasible set. Consequently we have
\begin{align}
f(S^{*})\geq &\frac{f(T_{2})+f(\widetilde{T}_{2})+f(T_{1}\cup\{\hat{e}_{2},\tilde{e}_{2}\})}{3}\notag\\
\geq &f(T_{1})+\frac{[f(T_{2})-f(T_{1})]+[f(\widetilde{T}_{2})-f(T_{1})]+[f(T_{1}\cup\{\hat{e}_{2},\tilde{e}_{2}\})-f(T_{1})]}{3} \notag\\
\geq & f(T_{1})+\tau_{2}\cdot \frac{c(T_{2}\setminus T_{1})+c(\widetilde{T}_{2}\setminus T_{1})+c(\hat{e}_{2})+c(\tilde{e}_{2})}{3}\notag\\
\geq & [f(\mathrm{OPT}^{\prime})-c(\mathrm{OPT}^{\prime})\cdot \tau_{1}]\cdot \mathbbm{1}_{c(T_{1})\leq \frac{1}{3}}+(c(T_{1})\cdot \tau_{1})\cdot \mathbbm{1}_{c(T_{1})\geq \frac{1}{3}}+\frac{2(1-c(T_{1}))}{3}\cdot \tau_{2}\notag\\
\geq & \Big(f(\mathrm{OPT}^{\prime})-c(\mathrm{OPT}^{\prime})\cdot \tau_{1}+\frac{4}{9}\tau_{2}\Big)\cdot \mathbbm{1}_{c(T_{1})\leq \frac{1}{3}}+\Big(\frac{\tau_{1}}{3}+\frac{4}{9}\tau_{2}\Big)\cdot \mathbbm{1}_{c(T_{1})\geq \frac{1}{3}}\label{combineqlast}.
\end{align}

\subparagraph*{Case $3$: Algorithm~\ref{streamingmain} stops without exceeding the budget.} In this case, we have
\begin{align}\label{lastcaseineq}
f(S^{*})\geq & f(\mathrm{OPT}^{\prime})-\sum_{e\in \mathrm{OPT}^{\prime}\setminus S^{*}}[f(S^{*}+e)-f(S^{*})] \notag\\
\geq & f(\mathrm{OPT}^{\prime})-\tau_{2}\cdot c(\mathrm{OPT}^{\prime}).	
\end{align}

Now we are ready to combine our analyses in the aforementioned three cases, \ie, inequalities (\ref{tau1vio}), (\ref{combine2}) and (\ref{combineqlast})-(\ref{lastcaseineq}), 
\begin{align}
f(S^{*})\geq &\min\Big\{\frac{\tau_{1}+\tau_{2}}{3}, f(\mathrm{OPT}^{\prime})-\tau_{1}\cdot c(\mathrm{OPT}^{\prime})+\frac{\tau_{2}}{3}, \notag
f(\mathrm{OPT}^{\prime})-\tau_{2}\cdot c(\mathrm{OPT}^{\prime})\Big\}	\\
\overset{(a)}{\geq} &\Big(\frac{6c(\mathrm{OPT}^{\prime})+1}{[3c(\mathrm{OPT}^{\prime})+1]^{2}}-\varepsilon \Big) \cdot f(\mathrm{OPT}^{\prime})\label{lowerboundoptprime}\\
\geq & \Big(\frac{7}{16}-\varepsilon \Big)f(\mathrm{OPT}) \cdot \mathbbm{1}_{\mathrm{OPT}\cap B=\varnothing}+\frac{16}{25}[f(\mathrm{OPT})-f(\mathrm{OPT}\cap B)]\cdot \mathbbm{1}_{\mathrm{OPT}\cap B\neq\varnothing}\label{lowerboundfstar}.
\end{align}
where $(a)$ holds with equality when
\begin{align}
\tau_{1}=\frac{3}{3c(\mathrm{OPT}^{\prime})+1}f(\mathrm{OPT}^{\prime})-O(\varepsilon)f(\mathrm{OPT})    
\end{align}
and 
\begin{align}
\tau_{2}=\frac{9c(\mathrm{OPT}^{\prime})}{[3c(\mathrm{OPT}^{\prime})+1]^{2}}f(\mathrm{OPT}^{\prime})-O(\varepsilon)f(\mathrm{OPT}).
\end{align}
Observe that the coefficient of $f(\mathrm{OPT}^{\prime})$ in (\ref{lowerboundoptprime}) decreases with respect to the weight of $\mathrm{OPT}^{\prime}$, thus (\ref{lowerboundfstar}) follows from the facts that $c(\mathrm{OPT}^{\prime})\leq 1-\frac{1}{2}\cdot \mathbbm{1}_{\mathrm{OPT}\cap B\neq \varnothing}$ and $f(\mathrm{OPT}^{\prime})\geq f(\mathrm{OPT})-f(\mathrm{OPT}\cap B)\cdot \mathbbm{1}_{\mathrm{OPT}\cap B\neq \varnothing}$.
\end{proof}

\subsection{Proof of Theorem \ref{budgettheorem}}

As a special case of Theorem \ref{psystemtheorem}, we have the following proposition.
\begin{proposition}\label{singleknappro}
There exists an $(4/11-\varepsilon)$-approximate algorithm for a single knapsack constraint, which performs $O(n\cdot\log\log n)$ queries.
\end{proposition}
\subsubsection{Approximation ratio}
\begin{proof}
We first show the following conclusion for set $S^{\prime}_{o}$, 
\begin{align}\label{lowerboundofguess}
f(S^{\prime}_{o})\geq \frac{4}{11}f(\mathrm{OPT})+ \Big(\frac{3}{11}-\delta \Big) f(\mathrm{OPT}\cap B).	
\end{align}	
Consider the iteration when
\begin{align*}
(1-\delta)f(\mathrm{OPT}\cap B)\leq \zeta=\hat{\zeta}\leq f(\mathrm{OPT}\cap B),    
\end{align*}
we claim that 
\begin{align*}
c(e_{\hat{\zeta}})\leq c(\mathrm{OPT}\cap B),
\end{align*}
since $\mathrm{OPT}\cap B$ is a candidate element when selecting element $e_{\hat{\zeta}}$. Moreover, $\mathrm{OPT}\setminus B$ is a feasible solution for the residual problem induced by $e_{\hat{\zeta}}$. Hence the following inequality holds for $S_{\hat{\zeta}}$, if we apply an $\beta$-approximation algorithm on the corresponding residual problem,
\begin{align}\label{residuallweround}
g_{\hat{\zeta}}(S_{\hat{\zeta}})\geq \beta \cdot g_{\hat{\zeta}}(\mathrm{OPT}\setminus B).	
\end{align}
Plugging the definition of the residual function $g_{\hat{\zeta}}(\cdot)$ into (\ref{residuallweround}), 
\begin{align}\label{resiineq}
f(S^{*})\geq f(S_{\hat{\zeta}}+e_{\hat{\zeta}})&\overset{(a)}{\geq} \beta \cdot f(\mathrm{OPT}\setminus B)	+(1-\beta)\cdot f(e_{\hat{\zeta}})\notag\\
&\overset{(b)}{\geq} \beta \cdot f(\mathrm{OPT})+(1-2\beta-\delta)\cdot f(\mathrm{OPT}\cap B),
\end{align}
where we use the fact that $g(\mathrm{OPT}\cap B)=f(\mathrm{OPT}\cap B+e_{\hat{\zeta}})-f(e_{\hat{\zeta}})\geq f(\mathrm{OPT}\cap B)-f(e_{\hat{\zeta}})$ in $(a)$; $(b)$ holds since $f(\mathrm{OPT}\setminus B)\geq f(\mathrm{OPT})-f(\mathrm{OPT}\cap B)$ and $f(e_{\hat{\zeta}})\geq \hat{\zeta} \geq (1-\delta)\cdot f(\mathrm{OPT}\cap B)$.  Recall that our Algorithm~\ref{mainalgorithmpsystemdknapsack} provides an approximation ratio of $\beta=\frac{4}{11}$, we can obtain~(\ref{lowerboundofguess}) by plugging $\beta=\frac{4}{11}$ into~(\ref{resiineq}).

Taken together with Lemma~\ref{propositionmpass}, we have
\begin{align}
f(S^{*})\geq &\min_{B}\max\Big\{\frac{7}{16} f(\mathrm{OPT}) \cdot \mathbbm{1}_{\mathrm{OPT}\cap B=\varnothing}+\frac{16}{25}[f(\mathrm{OPT})-f(\mathrm{OPT}\cap B)]\cdot \mathbbm{1}_{\mathrm{OPT}\cap B\neq\varnothing},\notag\\
&\;\;\;\;\;\;\;\;\;\;\;\;\;\;\;\;\;\;\;\;\frac{4}{11}f(\mathrm{OPT})+ \Big(\frac{3}{11}-\delta \Big) \cdot f(\mathrm{OPT}\cap B)\Big\}-O(\varepsilon) f(\mathrm{OPT})\label{deterdelta}\\
\geq&  \Big(\frac{7}{16}-\varepsilon \Big)\cdot f(\mathrm{OPT}) .
\end{align}
The proof is complete. 
\end{proof}

\subsubsection{Complexity of Algorithm \ref{guesslargest1}}
\paragraph{Offline time complexity.} Both Algorithm \ref{streamingmain} and Algorithm \ref{guesslargest1} require a constant approximations of $f(\mathrm{OPT})$, which can be obtained independently in $O(n\log\log n)$ time, for example, via similar treatments to the ADT algorithm. Notice that in Algorithm~\ref{streamingmain}, there are $O(\varepsilon^{-1})$ different values of $\lambda$ and the algorithm runs in $O(n)$ time for each fixed $\lambda$. Hence the total running time of Algorithm~\ref{streamingmain} is in the order of $O(n\cdot \max\{\varepsilon^{-1}, \log\log n\})$. Algorithm \ref{guesslargest1} can be accomplished within the same order of time, as it requires $O(\delta^{-1})$ calls to backtracking threshold Algorithm and $\delta=O(1)$.

\paragraph{Streaming setting.} In the streaming model, we run Algorithm \ref{streamingmain} and Algorithm \ref{guesslargest1} in parallel. Compared with offline algorithm, the main difference lies in the approach used to obtain a constant approximation of $f(\mathrm{OPT})$. Since $f(\mathrm{OPT})/\max_{e\in E}{f(e)}$ lies in the range of $[1,n]$, we can maintain $O(\log n/\varepsilon)$ copies of solutions in parallel for each possible approximation value of $f(\mathrm{OPT})$, which implies a total time complexity of $O((n\log n)/\varepsilon)$ and space complexity of $O((n\log n)/\varepsilon)$.

\section{A Nearly Linear Time $(1/2-\varepsilon)$-Approximate  Deterministic Algorithm for Binary Packing Constraints}\label{binarypacking}
We start with the formal definition about the residual problem with respect to a given set $T$.
\vspace{-0.3cm}
\begin{definition}[$T$-Residual Problem]
Let $f$ be a submodular function, its contracted function $f_{T}:2^{E\setminus T}\rightarrow \mathbbm{R}_{+}$ is given as $f_{T}(S)=f(S\cup T)-f(T)$. For the optimization problem $\max_{S\in \mathcal{I}}{f(S)}$, we define its $T$-residual problem as $\max_{S\in  \mathcal{I}_{T}}f_{T}(S)$, where the constraint set $\mathcal{I}_{T}=\{S|S\subseteq E\setminus T, S\cup T\in \mathcal{I}\}$.
\end{definition}

In several constrained submodular maximization problems~\cite{khuller1999budgeted,sviridenko2004note,kulik2009maximizing,ene2017nearly,badanidiyuru2014fast}, we are able to obtain a desirable approximation guarantee for the residual problem, by carefully choosing set $T$. For example, in the single knapsack constraint~\cite{sviridenko2004note}, $T$ represents the collection of three elements that have the largest marginal increments, while $T$ consists of elements with high costs for constant number of knapsack constraints~\cite{kulik2009maximizing}. However, directly searching set $T$ takes $O(n^{3})$ and $\Theta(n^{d})$ time respectively in the aforementioned two examples, which are computationally expensive.

Apart from the aforementioned straightforward approaches, we introduce the concept of \emph{shadow set}, and consider the residual problem with respect to $T^{\sharp}$, the shadow set of the target set $T$. The formal definition of shadow set is specified as follows.
\begin{definition}[$\alpha$-shadow set]~\label{definitionofsha} 
$T^{\sharp}$ is called $\alpha$-shadow set of $T\subseteq E$ iff $\mathrm{OPT}\setminus (T\cup T^{\sharp})$ is a feasible solution to $S^{\sharp}$-residual problem, while
\begin{align*}
f((\mathrm{OPT}\cup T^{\sharp})\setminus T)+\alpha \cdot f(T^{\sharp})\geq f(\mathrm{OPT}).
\end{align*}
\end{definition}
Definition~\ref{definitionofsha} states that replacing the optimal elements in $T$ with that in the $\alpha$-shadow set $T^{\sharp}$, will incur an additive loss that is no more than $\alpha \cdot f(T^{\sharp})$. 

\noindent\textbf{Algorithm overview.} We present our algorithm \ref{linearbinary} in Appendix \ref{appendixalgorithmmwu} and first introduce some necessary notations. Let $\mathrm{OPT}_{i}=\{o_{1},o_{2},\ldots,o_{i}\}$, $\Delta_{i}=f(\mathrm{OPT}_{i})-f(\mathrm{OPT}_{i-1})$ and
\begin{align*}
\ell_{\varepsilon}=\max\Big\{i\Big|\Delta_{i}\geq \frac{\varepsilon}{d}\cdot f(\mathrm{OPT})\Big\}.
\end{align*}
Without loss of generality, we assume that elements in $\mathrm{OPT}$ are in greedy ordering, \ie, $o_{i+1}=\argmax_{e \in \mathrm{OPT}}{\{f(\mathrm{OPT}_{i}+e)-f(\mathrm{OPT}_{i})\}}$. As shown in Algorithm~\ref{linearbinary}, we first construct $S^{\sharp}_{\varepsilon}$, a $(1+\varepsilon)$-shadow set of $\mathrm{OPT}_{i_{\varepsilon}}$, \ie, we select an element with comparable cost and similar marginal increment for each element in $\mathrm{OPT}_{i_{\varepsilon}}$.

We next consider the residual problem 
\begin{align*}
\max_{S\subseteq E\setminus S_{\varepsilon}^{\sharp}}\Big\{f(S\cup S_{\varepsilon}^{\sharp})\Big|S\subseteq E\setminus S_{\varepsilon}^{\sharp}, \mathbf{A}\mathbf{x}_{S\cup S_{\varepsilon}^{\sharp}}\leq \mathbf{b}\Big\},    
\end{align*}
for which we combine the MWU-based greedy algorithm \cite{azar2012efficient} with a threshold decreasing procedure on $E\setminus (E_{\Gamma}\cup S_{\varepsilon}^{\sharp})$, where $E_{\Gamma}=\{e\in E |\exists i\in \Gamma \mbox{ such that } c_{i}(e)\neq 0\}$ and 
\begin{align*}
\Gamma=\Big\{i\in [d]\Big|b_{i}-c_{i}(S^{\sharp})\leq W=\frac{2\log d}{\delta^{2}}\Big\}\subseteq [d].    
\end{align*}

The main ingredient of our algorithm is to construct $S_{\varepsilon}^{\sharp}$, the shadow set of $\mathrm{OPT}_{i_{\varepsilon}}$, by approximately guessing \emph{deterministically} in the value space, which enables us to establish a mapping between elements in $\mathrm{OPT}_{i_{\varepsilon}}$ and $S_{\varepsilon}^{\sharp}$. The analysis is in a similar spirit to the analysis of greedy algorithm under matroid constraint.

\begin{lemma}\label{shadowsetproposition}
$S_{\varepsilon}^{\sharp}$ is a $(1+\varepsilon)$-shadow set of $\mathrm{OPT}_{i_{\varepsilon}}$, \ie,
\begin{align*}
f(\mathrm{OPT})-f((\mathrm{OPT}\cup S^{\sharp})\setminus \mathrm{OPT}_{i_{\varepsilon}})\leq (1+\varepsilon) f(S^{\sharp}).
\end{align*}
\end{lemma}

\begin{proof}
For notational simplicity, we omit the subscript $\varepsilon$ in this proof. Let $S^{\sharp}=\{e^{\sharp}_{1}, e^{\sharp}_{2},\ldots,e^{\sharp}_{|S^{\sharp}|}\}$ and $S^{\sharp}_{i}=\{e^{\sharp}_{1}, e^{\sharp}_{2},\ldots,e^{\sharp}_{i}\}\;(i\in [|S^{\sharp}|])$, where $e^{\sharp}_{i}$ is the element selected at the $i$-th step of guessing. According to the definition of $\mathcal{G}_{\varepsilon}$ in Algorithm~\ref{linearbinary}, we know that the increment of $e^{\sharp}_{i}$ with respect to set $S^{\sharp}_{i-1}$ is similar as that of element $o^{\sharp}_{i}$, \ie, 
\begin{align}\label{incrementguess}
f(S^{\sharp}_{i})-f(S^{\sharp}_{i-1})&=f(S^{\sharp}_{i-1}+e^{\sharp}_{i})-f(S^{\sharp}_{i-1})\\
&\geq (1-\varepsilon)[f(S^{\sharp}_{i-1}+o_{i})-f(S^{\sharp}_{i-1})]-\frac{\varepsilon}{|\mathrm{OPT}_{i_{\varepsilon}}|}f(\mathrm{OPT}).
\end{align}
The increment of $o^{\natural}_{i}$ in (\ref{incrementguess}) can be lower bounded as
\begin{align}
f(S^{\sharp}_{i-1}+o_{i})-f(S^{\sharp}_{i-1})\overset{(a)}{\geq} &f(S^{\sharp}_{i-1} \cup (\mathrm{OPT}\setminus \mathrm{OPT}_{i-1}))-f(S^{\sharp}_{i-1}\cup (\mathrm{OPT}\setminus \mathrm{OPT}_{i}))\\
\overset{(b)}{\geq} &f(S^{\sharp}_{i-1} \cup (\mathrm{OPT}\setminus \mathrm{OPT}_{i-1}))-f(S^{\sharp}_{i}\cup (\mathrm{OPT}\setminus \mathrm{OPT}_{i}))
\end{align}
where in $(a)$ we use submodularity of $f$ and the fact that $S^{\sharp}_{i-1}\cup (\mathrm{OPT}\setminus \mathrm{OPT}_{i})+o_{i}=S^{\sharp}_{i-1} \cup (\mathrm{OPT}\setminus \mathrm{OPT}_{i-1})$. $(b)$ follows from monotonicity of $f$. Take summarization from $i=1$ to $|S^{\sharp}|$, we can obtain
\begin{align*}
f(S^{\sharp})=&\sum_{i=1}^{|S^{\sharp}|}{[f(S^{\sharp}_{i})-f(S^{\sharp}_{i-1})]}\\
\geq &(1-\varepsilon)\cdot\sum_{i=1}^{|S^{\sharp}|}{[f(S^{\sharp}_{i-1}+o_{i})-f(S^{\sharp}_{i-1})]}-\varepsilon f(\mathrm{OPT}) \\
\geq & (1-\varepsilon)\cdot\sum_{i=1}^{|S^{\sharp}|}{[ f(S^{\sharp}_{i-1} \cup (\mathrm{OPT}\setminus \mathrm{OPT}_{i-1}))-f(S^{\sharp}_{i}\cup (\mathrm{OPT}\setminus \mathrm{OPT}_{i}))]}-\varepsilon f(\mathrm{OPT}) \\
= & (1-\varepsilon)\cdot[f(\mathrm{OPT})-f((\mathrm{OPT}\cup S^{\natural})\setminus \mathrm{OPT}_{i_{\varepsilon}})]-\varepsilon f(\mathrm{OPT}).
\end{align*}
Rearranging the terms, the proof is complete.
\end{proof}

\paragraph{Remark.} Indeed we can further conclude that  $f(S_{\varepsilon}^{\sharp})\geq (1/2-\varepsilon) \cdot f(\mathrm{OPT}_{i_{\varepsilon}})$.

\begin{align*}
f(\mathrm{OPT}_{i_{\varepsilon}})-f(S^{\sharp})\leq& \sum_{e\in \mathrm{OPT}}{[f(S^{\sharp}+e)-f(S^{\sharp})]}\tag{submodularity}\\
=& \sum_{i=1}^{|S^{\sharp}|}{[f(S^{\sharp}_{i-1}+o_{i})-f(S^{\sharp}_{i-1})]}\\
\leq &\frac{1}{1-\varepsilon}\cdot\sum_{i=1}^{|S^{\sharp}|}{[f(S^{\sharp}_{i-1}+e^{\sharp}_{i})-f(S^{\sharp}_{i-1})]}\tag{selection rule of Algorithm \ref{linearbinary}}\\
=& \frac{f(S^{\sharp})}{1-\varepsilon}.
\end{align*}

\begin{proposition}\label{lemmaappguess} Algorithm \ref{linearbinary} returns a solution set $S_{o}$ in $O_{\varepsilon}(n\cdot \log \log n)$ time and $f(S_{o})\geq (\frac{1}{2}-\varepsilon)\cdot f(\mathrm{OPT})$.
\end{proposition}
\begin{proof}
See Appendix~\ref{appendixappguess}. 
\end{proof}

\noindent\textbf{Remark.} In general, a $\gamma$-approximate polynomial time algorithm for the $T$-residual problem implies a polynomial time $\min\{\gamma,\frac{1}{1+\beta}\}$-approximate algorithm. The proof is presented in Appendix~\ref{appendixclaim1}.

\section{Intersection of $p$-System and $d$-Knapsack Constraints}\label{psystemdknapsection}
In this section we consider the problem of maximizing a monotone submodular function under the intersection of a $p$ system constraint $\mathcal{I}_{p}$ and $d$ knapsack constraints, where $\mathcal{K}_{i}=\{S \subseteq E\;|\;c_{i}(S)\leq 1\}\;(\forall i\in [d])$ represents the $i$-th knapsack constraint. Element weights in the $i$-th dimension are specified by weight function $c_{i}(\cdot): 2^{E}\rightarrow \mathbbm{R}^{\geq 0}$.

\paragraph{Overview of the backtracking threshold algorithm.} As shown in Algorithm~\ref{backtrackingalgo1}, we eliminate elements with high cost that are collected by $B$, the set of large elements. Element $e\in E$ is called a \emph{large element} if $2c_{i}(e)>1$ holds for at least one index $i\in [d]$, otherwise we call it a small element. Among the remaining elements, those with marginal gain no less than $\Delta$ and profit density no less than the predetermined threshold $\theta$, will be added into the candidate set, as long as the newly constructed set is feasible. When the cost of the currently chosen element $e$ is larger than the residual budget, a new feasible solution $\widetilde{S}$ is constructed. Element $e$ and the element with largest total cost in set $S$ are firstly added into $\widetilde{S}$, we next add remaining elements in $S$ into $\widetilde{S}$ until exceeding the budget. We further use Algorithm~\ref{btadt} (presented in Appendix~\ref{appendixbtadt}), the combination of ADT and backtracking, to approximate $f(\mathrm{OPT})$.

\begin{algorithm}[H]
\label{backtrackingalgo1}
\small
\caption{Backtracking Threshold (BT) Algorithm ($\theta$, $\varepsilon$) (BT($\theta, \varepsilon$))}
\textbf{Initialization:} $S_{0}\leftarrow \{\argmax_{e\in E}f(e)\}$, $B\leftarrow \{e\in E\;|\;\exists i\in [d] \mbox{ such that } c_{i}(e)\geq 1/2\}$, $\Delta \leftarrow f(S_{0})$, $S\leftarrow \varnothing$\\
\While{$\Delta\geq \frac{\varepsilon f(S_{0})}{n}$}
{
\For{$e\in E \setminus B$}
{\If{$f_{S}(e)\geq \max\{\theta \cdot{\sum_{i=1}^{d}{c_{i}(e)}}, \Delta\}$ and $S+e \in \mathcal{I}_{p}$}
{\If{$S+e\in \mathcal{I}$}
{$S\leftarrow S+e$\\}
\Else{
$\widetilde{S}\leftarrow \{e,\;\argmax_{e\in S}{\sum_{i=1}^{d}{c_{i}}(e)}\}$\\
\For {$e\in S$}
{\If{$\widetilde{S}+e \in \mathcal{I}$}
{$\widetilde{S}\leftarrow \widetilde{S}+e$\\}
}
\textbf{break}
}
}
}
$\Delta \leftarrow (1-\varepsilon)\cdot\Delta$
}
\Return{$S^{*}\leftarrow \argmax_{T\in \{S, \tilde{S}, S_{0}\}}f(T)$}
\end{algorithm}

One may expect that removing large elements will incur a large loss in the objective value. However, the following two observations help to bound the loss. Firstly, note that the marginal gain of each element should be inversely related to $p$ and $d$ according to the desired approximation ratio, otherwise we can just return a singleton with objective larger than $(1/(p+\frac{7d}{4}+1)-\varepsilon)$ times optimum. Secondly, there are at most $d$ large elements in $\mathrm{OPT}$. On the other hand, the threshold selection procedure will be able to achieve higher objective value when the element costs are smaller (compared with the budget), with the additional help of recursively constructed set $\widetilde{S}$.

A simple but crucial consequence is the following upper bound on the number of large elements in $\mathrm{OPT}$. 
\begin{corollary}\label{numberofbigelebound}
There are at most $|\mathrm{OPT}\cap B|\leq d$ large elements in the optimal solution.	
\end{corollary}
\begin{proof}
Suppose that there are more than $d$ large elements in $\mathrm{OPT}$. From the pigeonhole principle we know that, there exists at least one index $i\in [d]$ such that $c_{i}(\mathrm{OPT})>1$. However, this contradicts the fact that $\mathrm{OPT}$ is a feasible solution set, the proof is complete. 
\end{proof}

\subsection{Monotone Submodular Maximization}\label{monotonepsystem}

Our algorithm for monotone objective is presented in Appendix~\ref{appendixalgpsystemdknapsack}, which computes the final solution by feeding a series of well-spaced parameters that are related to the output of Algorithm~\ref{btadt}. The proof of Theorem \ref{psystemtheorem} is presented in Appendix \ref{appendixlemmapsystem}.

Let $e^{\flat}$ be the element that is not added to $S$ due to violation of some knapsack constraints in Algorithm~\ref{backtrackingalgo1}. As element $e^{\flat}$ may not exist, we divide the analysis of backtracking threshold algorithm into two cases in Proposition~\ref{propositioneexists} and \ref{tildeeexist}, based on the existence of $e^{\flat}$.
\begin{proposition}
\label{propositioneexists}
If element $e^{\flat}$ does not exist,
\begin{align*}
f(S^{*}) \geq \frac{f(\mathrm{OPT})-\theta\cdot (d-|\mathrm{OPT}\cap B|/2)}{p+|\mathrm{OPT}\cap B|+1}-(p\varepsilon+\varepsilon)\cdot f(\mathrm{OPT}).    
\end{align*}
\end{proposition}
\begin{proof}
We partition the optimal solution set as 
\begin{align*}
\mathrm{OPT}=\mathrm{OPT}_{1}\cup \mathrm{OPT}_{2}\cup (\mathrm{OPT}\cap B),    
\end{align*}
where $\mathrm{OPT}_{1}=\{e\in \mathrm{OPT}\setminus B|f_{S^{\flat}}(e)< \theta \cdot \sum\nolimits_{j=1}^{d}{c_{i}(e)}\}$ represents the set of small elements in $\mathrm{OPT}$, whose profit density with respect to set $S^{\flat}$ is less than $\theta$, $\mathrm{OPT}_{2}=\mathrm{OPT}\setminus (B\cup \mathrm{OPT}_{1})$ denotes the remaining small elements in $\mathrm{OPT}$. Based on this partition, we are able to lower bound $f(S)$ in the following manner,
\begin{align}
f(\mathrm{OPT})-f(S) \overset{(a)}{\leq}& f(S \cup \mathrm{OPT})-f(S) \tag{monotonicity}\\
= &[f(S\cup (\mathrm{OPT}\cap B))-f(S)]+[f(S\cup (\mathrm{OPT}\cap B)\cup \mathrm{OPT}_{1})-f(S\cup (\mathrm{OPT}\cap B))]\notag\\
& + [f(S\cup \mathrm{OPT})-f(S\cup (\mathrm{OPT}\cap B)\cup \mathrm{OPT}_{1})] \notag\\ 
\overset{(b)}{\leq} & 	\underbrace{[f[S\cup (\mathrm{OPT}\cap B)]-f(S)]}_{\Sigma_{1}}+\underbrace{[f(S\cup \mathrm{OPT}_{1})-f(S)]}_{\Sigma_{2}}+\underbrace{[f(S\cup \mathrm{OPT}_{2})-f(S)]}_{\Sigma_{3}} \label{3upperbound}
\end{align}
where $(b)$ follows from submodularity of $f(\cdot)$.

In the following, we provide upper bounds on $\Sigma_{1}, \Sigma_{2}, \Sigma_{3}$ respectively. Firstly, a direct consequence of submodularity and the definition of $S^{*}$ is,
\begin{align}\label{rb}
\Sigma_{1}=f(S\cup (\mathrm{OPT}\cap B))-f(S)\leq \sum_{e\in \mathrm{OPT}\cap B}{f(e)}\leq|\mathrm{OPT}\cap B|\cdot f(S^{*}).
\end{align}
As for the second term $\Sigma_{2}$,
\begin{align}\label{r1}
\Sigma_{2}=f(S\cup \mathrm{OPT}_{1})-f(S)=&\sum_{e\in \mathrm{OPT}_{1}}{[f(S+e)-f(S)]}\overset{(a)}{\leq} \theta \cdot \sum_{j=1}^{d}{c_{j}(\mathrm{OPT}_{1})}\notag\\
\overset{(b)}{\leq} & \theta \cdot \Big(d-\sum_{j=1}^{d}{c_{j}(\mathrm{OPT}\cap B)} \Big)\notag\\
\overset{(c)}{\leq} & \theta \cdot \Big(d-\frac{|\mathrm{OPT}\cap B|}{2}\Big),
\end{align}
where $(a)$ is based on the definition of $\mathrm{OPT}_{1}$, which indicates that the profit density of elements in $\mathrm{OPT}_{1}$ is less than $\theta$. The correctness of  $(b)$ follows from the fact that $c_{j}(\mathrm{OPT}_{1})\leq c_{j}(\mathrm{OPT})-c_{j}(\mathrm{OPT}\cap B)\leq 1-c_{j}(\mathrm{OPT}\cap B)$ $(\forall j\in [d])$. $(c)$ holds because the total cost of each large element is no less than $1/2$.

We next introduce Proposition~\ref{psystemlemma}, whose proof follows from the analysis of greedy algorithm for monotone objective and $p$-system constraint~\cite{badanidiyuru2014fast, calinescu2011maximizing}. We include its proof in Appendix~\ref{appendixpsystem} for completeness. 
\begin{proposition}\label{psystemlemma}
$f(S\cup \mathrm{OPT}_{2})-f(S)\leq [p+(p+1)\varepsilon]\cdot f(S)$.
\end{proposition}
Assuming Proposition~\ref{psystemlemma}, we are able to lower bound the objective value of $S$.
More specifically, by plugging inequalities (\ref{rb})--(\ref{r1}) and Proposition \ref{psystemlemma} into (\ref{3upperbound}), we have
\begin{align}
f(S)\geq &f(\mathrm{OPT})-(\Sigma_{1}+\Sigma_{2}+\Sigma_{3})\notag\\
\geq & f(\mathrm{OPT})-|\mathrm{OPT}\cap B|\cdot f(S^{*})-	\theta \cdot(d-|\mathrm{OPT}\cap B|/2)-(p+O(\varepsilon))f(S).
\end{align}
Rearranging the terms,
\begin{align}
(p+1+|\mathrm{OPT}\cap B|+O(\varepsilon))\cdot f(S^{*})\geq &(p+1+O(\varepsilon))\cdot f(S)+|\mathrm{OPT}\cap B|\cdot f(S^{*})\notag\\
\geq &f(\mathrm{OPT})-	\theta \cdot(d-|\mathrm{OPT}\cap B|/2).
\end{align}
The proof is complete.
\end{proof}
Before providing the lower bound of $f(S^{*})$ for the case when $e^{\flat}$ exists, we first show the feasibility of all the candidate solutions involved. Let $S^{\flat}$ be the value of candidate solution $S$ before considering $e^{\flat}$, then there exists $i\in [d]$ such that $c_{i}(S^{\flat}+e^{\flat})>1$. We use $\hat{e}=\argmax_{e\in S^{\flat}}{\sum\nolimits_{i=1}^{d}{c_{i}(e)}}$ to represent the element with the largest total cost in $S^{\flat}$ and let $\widetilde{S}^{(1)}=\{e^{\flat},\hat{e}\}$, $\widetilde{S}^{(2)}=\widetilde{S}\setminus \widetilde{S}^{(1)}$. 
\begin{proposition}\label{factfeasible}
Both $\widetilde{S}$ and $\widetilde{S}^{(1)}$ are feasible solution sets.	
\end{proposition}
\begin{proof}
Note that $\widetilde{S}$ is initialized to $\widetilde{S}^{(1)}$ and is always feasible after each update, it suffices to show that $\widetilde{S}^{(1)}\in \mathcal{I}$. We first remark that $S^{\flat}\cup\{e^{\flat}\} \in \mathcal{I}_{p}$, since $e^{\flat}$ satisfies all conditions required in line $4$ of Algorithm~\ref{backtrackingalgo1}. According to the \emph{down-closed} property of $\mathcal{I}_{p}$ and the fact that $\widetilde{S}^{(2)}\subseteq S^{\flat}\cup\{e^{\flat}\}$, we conclude that $\widetilde{S}^{(2)}\in \mathcal{I}_{p}$. It remains to show that $\widetilde{S}^{(2)}$ belongs to $\cap_{i=1}^{d}{\mathcal{K}_{i}}$. Recall that both $e^{\flat}$ and $\hat{e}$ are small elements, then we have 
\begin{align*}
c_{j}(\widetilde{S}^{(2)})\leq c_{j}(e^{\flat})+c_{j}(\hat{e})\leq 1   
\end{align*}
 for $\forall j\in [d]$, which implies that $\widetilde{S}^{(2)}\in \cap_{i=1}^{d}{\mathcal{K}_{i}}$. The proof is complete.
\end{proof}

\begin{proposition}\label{tildeeexist} 
$3f(S^{*})\geq 2\theta$ holds when element $e^{\flat}$ exists.
\end{proposition}
\begin{proof}
It is standard to assume that every singleton in $E$ is feasible, otherwise we can apply our algorithm on the ground set consisting of feasible singletons. Without loss of generality we assume that $S^{\flat}=\{e_{1}, e_{2},\ldots,e_{|S^{\flat}|}\}$, where $e_{i}$ denotes the $i$-th element added into $S^{\flat}$. Since $\widetilde{S}^{(2)}$ is also a subset of $S^{\flat}$, we denote it as $\widetilde{S}^{(2)}=\{e_{i_1},e_{i_2},\ldots,e_{i_{|\widetilde{S}^{(2)}|}}\} $, where $i_{\ell}\in [|S^{\flat}|]$ for $\forall \ell\leq |\widetilde{S}^{(2)}|\leq |S^{\flat}|$. We further let $S^{\flat}_{i}=\{e_{1},e_{2},\ldots,e_{i}\}\;(i\in [|S^{\flat}|])$ be the first $i$ elements in $S^{\flat}$, and $\widetilde{S}^{(2)}_{i}$ is defined in an analogous manner. 

According to the density threshold rule, we have 
\begin{align}\label{pdthresholdrule}
f(S^{\flat}_{i+1})-f(S^{\flat}_{i})\geq \theta \cdot \sum_{j=1}^{d}{c_{j}(e_{i+1})}.
\end{align}
The objective value of $\widetilde{S}^{(2)}$ can be lower bounded as
\begin{align}
f(\widetilde{S}^{(2)})&=\sum_{j=1}^{|\widetilde{S}^{(2)}|}{[f(\widetilde{S}^{(2)}_{j})-f(\widetilde{S}^{(2)}_{j-1})]}\geq \sum_{j=1}^{|\widetilde{S}^{(2)}|}{[f(S^{\flat}_{i_{j}})-f(S^{\flat}_{i_{j}-1})]}\tag{submodularity}\\
&\geq \theta \cdot \sum_{j=1}^{|\widetilde{S}^{(2)}|}{\sum_{t=1}^{d}{c_{t}(e_{i_j})}}=\theta \cdot \sum_{t=1}^{d}{c_{j}(\widetilde{S}^{(2)})},\label{lowboundset2}
\end{align}
where the last inequality is due to (\ref{pdthresholdrule}). Similarly we have
\begin{align}\label{pdthresholdineq}
f(\widetilde{S})\geq \theta \cdot \sum_{t=1}^{d}{c_{j}(\widetilde{S})}.
\end{align}
We next claim that $S^{\flat}\setminus (\widetilde{S}^{(2)}+\hat{e})$ is non-empty, otherwise $\widetilde{S}$ will be equal to $S^{\flat}+e^{\flat}$ and this contradicts the fact that $\widetilde{S}$ is a feasible solution. Hence there exists at least one element $\bar{e}\in S^{\flat}\setminus (\widetilde{S}^{(2)}+\hat{e})$. We further note that there exists at least one index $i^{\dag}$, such that $c_{i^{\dag}}(\widetilde{S}+\bar{e})>1$. Otherwise we have $\widetilde{S}+\bar{e}\in \cap_{i=1}^{d}{\mathcal{K}_{i}}$. Moreover, the fact that $\widetilde{S}+\bar{e}\subseteq S^{\flat}+e^{\flat}$ and $S^{\flat}+e^{\flat}\in \mathcal{I}_{p}$ implies that $\widetilde{S}+\bar{e}$ also belongs to $ \mathcal{I}_{p}$. As a consequence, $\bar{e}$ will be added into $\tilde{S}_{\tilde{e}}$, this contradicts the fact that $\bar{e}\in S^{\flat}\setminus (\widetilde{S}^{(2)}+\hat{e})$. Therefore
\begin{align}\label{lowerboundcost2}
\sum\nolimits_{j=1}^{d}{c_{j}(\widetilde{S}+\bar{e})}	\geq c_{i^{\dag}}(\widetilde{S}+\bar{e})>1.
\end{align}
By combining (\ref{lowboundset2}) with (\ref{lowerboundcost2}), and interchanging the order of the summation, we know that 
\begin{align}\label{bound1}
f(\widetilde{S})\geq \theta \cdot \sum_{j=1}^{d}{c_{j}(\widetilde{S})}\geq \theta \cdot \Big(1-\sum_{j=1}^{d}{c_{j}(\bar{e})}	\Big)\geq \theta \cdot \Big(1- \sum_{j=1}^{d}{c_{j}(\hat{e})}\Big),
\end{align}
where the first and second inequality follow from (\ref{pdthresholdineq}) and (\ref{lowerboundcost2}) respectively, the last inequality is based on the definition of $\hat{e}$. On the other hand, using similar arguments to (\ref{lowboundset2}) and the monotonicity of $f$, we have
\begin{align}\label{bound2}
f(\widetilde{S})\geq f(\widetilde{S}^{(1)})	\geq \theta \cdot\Big(\sum\nolimits_{j=1}^{d}{c_{j}(e^{\flat})}+ \sum\nolimits_{j=1}^{d}{c_{j}(\hat{e})}\Big).
\end{align}
Moreover, 
\begin{align}
f(S^{\flat})\geq \theta \cdot \Big(\sum_{e\in S }{\sum\nolimits_{j=1}^{d}{c_{j}(e)}} \Big)
\geq \theta \cdot \Big(1-\sum\nolimits_{j=1}^{d}{c_{j}(e^{\flat})}\Big),\label{bound3}
\end{align}
where the last inequality holds because the total costs of $S^{\flat}+e^{\flat}$ in all the $d$ dimensions is larger than $1$, since $S^{\flat}+e^{\flat}$ belongs to $\mathcal{I}_{p}$ but $S^{\flat}+e^{\flat}\notin \mathcal{I}$. Combining (\ref{bound1})--(\ref{bound3}), we are able to derive the following lower bound on the quality of output set $S^{*}$,
\begin{align}\label{lowercase1}
f(S^{*})= &\max{\{f(\widetilde{S}),f(S^{\flat})\}} \geq \frac{f(S^{\flat})+2f(\widetilde{S})}{3}\notag\\
\geq &\frac{1}{3} \Big[\theta \cdot \Big(1-\sum_{j=1}^{d}{c_{j}(\hat{e})}\Big)+\theta \cdot \Big(\sum_{j=1}^{d}{c_{j}(\hat{e})}+ \sum_{j=1}^{d}{c_{j}(e^{\flat})}\Big)+\theta \cdot \Big(1-\sum_{j=1}^{d}{c_{j}(e^{\flat})}\Big) \Big]\notag\\
=& \frac{2}{3}\theta.
\end{align}
The proof is complete.
\end{proof}

\subsection{Non-Monotone Submodular Maximization}\label{nonmonotoneicml}
We extend our algorithm to non-monotone submodular functions and present an algorithm that achieves a better approximation ratio than that in~\cite{mirzasoleiman2016fast}. See table~\ref{nonmonotonetable} for a detailed summary. Mirzasoleiman et al.~\cite{mirzasoleiman2016fast} designed a fast algorithm that achieves an approximation ratio of  $(\frac{p/(p+1)}{2p+2d+1}-\varepsilon)$. The algorithm is a combination of two paradigms---algorithm for maximizing monotone submodular function under $p$-system+$d$-knapsack constraint~\cite{badanidiyuru2014fast}, and algorithm for maximizing a non-monotone submodular function under $p$-system constraint~\cite{gupta2010constrained}. Hence it is natural to expect a better performance guarantee via our techniques developed for monotone submodular functions in Section~\ref{monotonepsystem}.  

\begin{table}[h]
\centering
\begin{tabular}{|c|c|c|c|}
     \hline
     \multicolumn{4}{|c|}{Comparisons Between Existing Algorithms}\\
     \hline\hline
     \textbf{Reference} & \textbf{Constraint} & \textbf{Approximation Ratio} & \textbf{Query Complexity} \\
     \hline
 \cite{feldman2011nonmonotone,vondrak2011submodular}  & $1$-matroid+$d$-knapsack  & $1/e-\varepsilon$  & $\poly(n)\cdot \exp(d,\varepsilon^{-1})$  \\ \hline
 \cite{vondrak2011submodular}  & $p$-matroid+$d$-knapsack & $0.19/p-\varepsilon$ & $\poly(n)\cdot \exp(p,d,\varepsilon^{-1})$ \\ \hline
\cite{mirzasoleiman2016fast} & $p$-system+$d$-knapsack & $\frac{p/(p+1)}{2p+2d+1}-\varepsilon$ & $O(nrp\cdot \log n/\varepsilon)$ \\ \hline
  This Paper & $p$-system+$d$-knapsack & $\frac{p/(p+1)}{2p+\frac{7}{4}d+1}-\varepsilon$  & $O(nrp\cdot \max\{\varepsilon^{-1},\log\log n\})$ \\
    \hline
    \end{tabular}
\caption{Non-monotone submodular maximizing under p-system and d-knapsack constraints.} 
\label{nonmonotonetable}
\end{table}

For our improved approximation ratio, we mainly highlight its lower dependence on the number of knapsack constraints. We achieve the performance guarantee via the following treatments.
\begin{itemize}
\item Different from the \emph{Greedy with Density Threshold} (GDT) algorithm, \ie, Algorithm $1$ of~\cite{mirzasoleiman2016fast}, we employ our backtracking algorithm to select feasible solution sets with the desired objective value. 
\item Similar to the approach in~\cite{gupta2010constrained}, the \emph{Iterated Greedy with Density Threshold} (IGDT) algorithm is introduced to overcome the non-monotonicity~\cite{mirzasoleiman2016fast}, in which GDT appears as the key subroutine for solution selection over various ground sets. Again we utilize backtracking threshold algorithm to replace the GDT algorithm in~\cite{mirzasoleiman2016fast}.
\item Unlike the FANTOM algorithm, \ie, Algorithm $3$ of~\cite{mirzasoleiman2016fast}, we are able to reduce query complexity in the non-monotone case, via a similar fashion as in our ADT algorithm.
\end{itemize}

In the following, we provide the key arguments for our analysis, and omit details that virtually follow our proof in Section~\ref{psystemdknapsection}.

\subsubsection{Performance Analysis}

\paragraph{Backtracking Algorithm in non-monotone case.}
We have the following conclusion about the quality of the set returned by backtracking algorithm, when the submodular function is non-monotone. The proof of Proposition~\ref{monotoneextendnon} is almost identical as that of Proposition~\ref{propositioneexists} and \ref{tildeeexist}, where the main difference is that, the benchmark quantity is changed to $f(S\cup \mathrm{OPT})$. 

\begin{proposition} \label{monotoneextendnon}
Algorithm $BT(\theta,\varepsilon)$ returns set $S\in \mathcal{I}$ such that  for any $T\in \mathcal{I}$,
\begin{align}\label{nonmonotoneminineq}
f(S)\geq \min\Big\{\frac{2}{3}\theta, \frac{f(S\cup T)-\theta ( d-|T\cap B|/2)}{p+|T\cap B|+1}-(p\varepsilon+\varepsilon)\cdot f(\mathrm{OPT}) \Big\},
\end{align}
where $B$ refers to the set of large elements.
\end{proposition}

\paragraph{IGDT with BT as a subroutine.} Following notations in \cite{mirzasoleiman2016fast}, we have the following proposition with respect to the new IGDT algorithm.
\begin{proposition}\label{nonmonotonelemma}
In each iteration of the IGDT algorithm with BT as a subroutine, we have
\begin{align}
\mathbb{E}[\max_{i\in [p+1]}\{f(S^{\prime}_{i}),f(S_{i})\}]\geq \min \Big\{\frac{2}{3}\theta,\;& \frac{p}{(p+1)(2p+1)+\sum_{i=1}^{p+1}{|\mathrm{OPT}_{i}\cap B|}}\cdot f(\mathrm{OPT}) \notag\\
&-\frac{(p+1)d-(\sum_{i=1}^{p+1}{|\mathrm{OPT}_{i}\cap B|})/2}{(p+1)(2p+1)+\sum_{i=1}^{p+1}{|\mathrm{OPT}_{i}\cap B|}}\cdot \theta\Big\}	
\end{align}	
\end{proposition}

\begin{proof}
In the rest of this proof, we let $O_{i}=O\setminus(O\cap \cup_{j=1}^{i-1}{S_{j}})$. Compared with \cite{mirzasoleiman2016fast}, we present a simpler and cleaner proof on the lower bound of $\sum_{i=1}^{p+1}{f(S_{i}\cup O_{i})}$. The key argument in our proof is the following telescoping sum, 
\begin{align}
f(S_{i}\cup O_{i})\geq f(O_{i+1})+[f((\cup_{j=i}^{p+1}{S_{j}})\cup O_{i})-f((\cup_{j=i+1}^{p+1}{S_{j}})\cup O_{i+1})].
\end{align}
This follows from submodularity of $f$, together with the facts that 
\begin{align}
(S_{i}\cup O_{i})\setminus O_{i+1}=[(\cup_{j=i}^{p+1}{S_{j}})\cup O_{i}]\setminus [(\cup_{j=i+1}^{p+1}{S_{j}})\cup O_{i+1}]=S_{i},
\end{align}
and $O_{i+1}\subseteq (\cup_{j=i+1}^{p+1}{S_{j}})\cup O_{i+1}$. Take summation over $i\in [p+1]$, we can obtain that
\begin{align}
\sum_{i=1}^{p+1}{f(S_{i}\cup O_{i})}	\geq f((\cup_{j=1}^{p+1}{S_{j}})\cup O_{i})+\sum_{i=2}^{p+1}{f(O_{i})}\geq \sum_{i=2}^{p+1}{f(O_{i})},
\end{align}
where the second inequality follows from the non-negativity of $f((\cup_{j=1}^{p+1}{S_{j}})\cup O_{i})$. 

Furthermore, note that
\begin{align}
p(p+1)\cdot \mathbb{E}[\max\{f(S^{\prime}_{i}),f(S_{i})\}]\geq &p(p+1)\cdot \mathbb{E}[f(S^{\prime}_{i})]=2\sum_{i=1}^{p+1}{(p+1-i)\cdot \mathbb{E}[f(S^{\prime}_{i})]}\notag\\
\geq & \sum_{i=1}^{p+1}{(p+1-i)\cdot f(S_{i}\cap O)},
\end{align}
where the last inequality follows from the approximation guarantee of the double greedy algorithm~\cite{buchbinder2012tight}. On the other hand,
\begin{align}
&\Big[(p+1)^{2}+\sum_{i=1}^{p+1}{|O_{i}\cap B|}\Big]\cdot \mathbb{E}[\max\{f(S^{\prime}_{i}),f(S_{i})\}]\notag\\
\geq & \sum_{i=1}^{p+1}{[p+|O_{i}\cap B|+1]\cdot f(S_{i})}\notag\\
\ge & \sum_{i=1}^{p+1}{f(S_{i}\cup O_{i})}-\Big[(pd+d)\theta-\frac{1}{2}\sum_{i=1}^{p+1}{|O_{i}\cap B|\cdot \theta}+O(\varepsilon)OPT\Big] \tag{Proposition \ref{monotoneextendnon}}
\end{align}
Combining together, we have
\begin{align}
&\Big[(p+1)(2p+1)+\sum_{i=1}^{p+1}{|O_{i}\cap B|}\Big]\cdot \mathbb{E}[\max\{f(S^{\prime}_{i}),f(S_{i})\}]\notag\\ 
\geq &\sum_{i=1}^{p+1}{(p+1-i)f(S_{i}\cap O)}+\sum_{i=1}^{p+1}{f(S_{i}\cup O_{i})}-((p+1)d\theta-\frac{1}{2}\sum_{i=1}^{p+1}{|O_{i}\cap B|\cdot \theta}+O(\varepsilon)\cdot \mathrm{OPT})\notag\\
\geq & [p-O(\varepsilon)]\cdot \mathrm{OPT}-[(p+1)d-\frac{1}{2}\sum_{i=1}^{p+1}{|O_{i}\cap B |}]\cdot\theta.
\end{align}
%
The proof is complete, since $\max_{i\in [p+1]}\{f(S^{\prime}_{i}),f(S_{i})\}\geq 2\theta/3$ directly follows from inequality (\ref{nonmonotoneminineq}).	
\end{proof}

\paragraph{Proof of Theorem \ref{nonmonotoneknappsys}.}
The details of our ADT algorithm in the non-monotone case is presented in Appendix~\ref{algononmonotonepd}, which is similar to Algorithm~\ref{btadt}. The proof of Theorem \ref{nonmonotoneknappsys} is similar as that for Theorem \ref{psystemtheorem}, except that the optimal threshold $\theta^{*}$ is specified by the following equation, 
\begin{align*}
\theta^{*}&=\frac{6p}{4(p+1)(2p+1)+\sum_{i=1}^{p+1}{|\mathrm{OPT}_{i}\cap B|}+(6p+6)d}\cdot f(\mathrm{OPT})\\
&\in \Big[\frac{6p}{(p+1)(8p+7d+4)}\cdot f(\mathrm{OPT}), f(\mathrm{OPT})\Big]. \tag{ $0\leq |\mathrm{OPT}_{i}\cap B|\leq d$}
\end{align*}




\bibliographystyle{plain}	
\bibliography{paper}

\appendix
\section{Supplementary Materials of Section~\ref{knapsacksec}}
\subsection{Comparison of ADT with algorithm in~\cite{huang2018multi}}\label{adtcompare}

The main subroutine in~\cite{huang2018multi} requires an $(1-\varepsilon)$-approximation of $f(\mathrm{OPT})$, which is obtained by applying a binary search procedure on $O(\varepsilon^{-1}\log n)$ predefined well-spaced guesses. In each iteration of the binary search, the main subroutine procedure is applied, which requires $O(\varepsilon^{-1})$ passes on the elements. However, the preprocessing phase of our ADT uses thresholds decreasing at a speed that is related to the numerical output in last iteration, and update the estimations according to objective values of $\lfloor\log (\bar{w}_{q}/ \underline{w}_{q})/\log (1+\alpha_{q}) \rfloor$ sets. ADT only makes a single pass on the elements to get the result for each fixed threshold. More importantly, while ADT maintains an interval $[\underline{w}_{i}, \bar{w}_{i+1}]$ that contains $f(\mathrm{OPT})$ and shrinks at each iteration, the algorithm in~\cite{huang2018multi} is not able to obtain estimates of $f(\mathrm{OPT})$ until the whole binary search procedure terminates, as it requires the value of all the historic midpoints during the binary search.

\section{Supplementary Materials of Section~\ref{binarypacking}}

\subsection{Details of Algorithm~\ref{linearbinary}}\label{appendixalgorithmmwu}
\begin{algorithm}[H]
\label{linearbinary}
\small
    \caption{An $O_{\varepsilon}(n\log \log n)$ Time Deterministic Algorithm for $\mathbf{A}\in \{0,1\}^{O(1)\times n}$}
Construct a $(\frac{1}{2}-\varepsilon, 1+\varepsilon)$-shadow set of $\mathrm{OPT}_{i_{\varepsilon}}$, denoted by $S^{\sharp}$, by (1) guessing the marginal increments of elements in $\mathrm{OPT}_{i_{\varepsilon}}$ via set $\mathcal{G}_{\varepsilon}=\{0, \frac{\varepsilon}{|O_{i_{\varepsilon}}|} f(\mathrm{OPT}), \frac{2\varepsilon}{|O_{i_{\varepsilon}}|} f(\mathrm{OPT}),\cdots, f(\mathrm{OPT})\}$; (2) guessing the cost vector of elements in $O_{i_{\varepsilon}}$, which belongs to $\{0,1\}^{d}$. \\
$b^{\prime}_{i}\leftarrow b_{i}-c_{i}(S^{\sharp})\;(\forall i\in [d])$\\
$w_{i}\leftarrow \frac{1}{b^{\prime}_{i}}\;(\forall i\in [d]$)\\
$\lambda, \bar{\lambda},\underline{\lambda}\leftarrow $ Upper and lower bounds on the ratio of marginal gain and weight, $T\leftarrow \varnothing$\\
\While{$\lambda\geq \underline{\lambda}$}
{
\For{$e\in E\setminus (E_{\Gamma}\cup S^{\sharp}\cup T)$}
{
\If{$\sum_{i=1}^{d}{b^{\prime}_{i}w_{i}}\leq 1+\delta W+(\delta W)^{2}$ and $f_{T\cup S^{\sharp}}(e) \geq \lambda\cdot \sum_{i=1}^{d}{w_{i}c_{i}(e)}$}
{ 
$w_{i}\leftarrow \Big[1+\frac{\delta W c_{i}(e)}{b^{\prime}_{i}}+\Big(\frac{\delta W c_{i}(e)}{b^{\prime}_{i}}\Big)^{2}\Big]\cdot w_{i}\;(\forall i\in [d])$\\
$T\leftarrow T+e$\\
}
}
$\lambda \leftarrow (1-\varepsilon)\cdot \lambda$
}
\textbf{Return} $S_{o}=S^{\sharp}_{\varepsilon}\cup T$
\end{algorithm}

\subsection{Proof of Proposition~\ref{lemmaappguess}}\label{appendixappguess}
\begin{proof}
We first note that $|\mathrm{OPT}\cap E_{\Gamma}|\leq dW$, since for each $i\in \Gamma$, there are at most $W$ elements in $\mathrm{OPT}\cap E_{\Gamma}$ whose cost in the $i$-th dimension is non-zero and $|\Gamma|\leq d$. We next claim the following proposition about the performance of Proposition~\ref{mwuproposition}.

\begin{proposition}\label{mwuproposition}
Set $T$ is an $(1-1/e-\delta)$-approximate solution for the $S^{\sharp}$--residual problem, \ie,
\begin{align} 
f(T\cup S^{\sharp})-f(S^{\sharp})\geq (1-1/e-O(\delta) )[f(\mathrm{OPT}\cup S^{\sharp}\setminus (\mathrm{OPT}_{i_{\varepsilon}}\cup E_{\gamma}))-f(S^{\sharp})].	
\end{align}
\end{proposition}

Assuming the correctness of Proposition~\ref{mwuproposition}, we are ready to finish the proof. According to Proposition~\ref{shadowsetproposition}, we have
\begin{align}
f(\mathrm{OPT}\setminus E_{\gamma})-f((\mathrm{OPT}\cup S^{\sharp})\setminus (\mathrm{OPT}_{i_{\varepsilon}}\cup E_{\gamma}))\leq f(S^{\sharp}).
\end{align}
Combining with Proposition \ref{mwuproposition} in which we let $\delta=\frac{1}{2}-\frac{1}{e}$, we know that
\begin{align}
f(T\cup S^{\sharp})\geq \frac{f(\mathrm{OPT}\setminus E_{\Gamma})}{2}\geq \Big(\frac{1}{2}-\varepsilon \Big)f(\mathrm{OPT}),
\end{align}
which follows from the definition of $E_{\Gamma}$, as removing each element in $O\cap E_{\Gamma}$ will incur a loss no more than $\frac{\varepsilon }{dW}f(\mathrm{OPT})$ and $|\mathrm{OPT}\cap E_{\Gamma}|\leq dW$.	

\end{proof}

\begin{proofof}{Proposition~\ref{mwuproposition}} 
In this proof we shall depart from the previous notation, for any set $S\subseteq E$, we let $S^{\prime}=S\setminus (S^{\sharp}\cup E_{\Gamma})$. Suppose that element $e^{(t+1)}$ is selected at the $(t+1)$-th iteration of the \emph{while} loop, then for $\forall e\in E^{\prime}\setminus T^{(t)}$, we have
\begin{align}\label{mwuselectionrule}
f_{T^{(t)}\cup S^{\sharp}}(e)\leq \frac{\sum_{i=1}^{d}{w^{(t)}_{i}c_{i}(e)}}{\sum_{i=1}^{d}{w^{(t)}_{i}c_{i}(e^{(t+1)})}}	f_{T^{(t)}\cup S^{\sharp}}(e^{(t+1)}),
\end{align}
which implies that
\begin{align*}
\sum_{e\in \mathrm{OPT}^{\prime}}{f_{T^{(t)}\cup S^{\sharp}}(e)	}\leq \frac{\sum_{i=1}^{d}{w^{(t)}_{i}c_{i}(\mathrm{OPT}^{\prime})}}{\sum_{i=1}^{d}{w^{(t)}_{i}c_{i}(e^{(t+1)})}}	f_{T^{(t)}\cup S^{\sharp}}(e^{(t+1)}).
\end{align*}
For the LHS, based on the submodularity of $f$, we have 
\begin{align}\label{submwu}
\sum_{e\in \mathrm{OPT}^{\prime}}{f_{T^{(t)}\cup S^{\sharp}}(e)}\geq f(\mathrm{OPT}^{\prime} \cup S^{\sharp})-f(T^{(t)} \cup S^{\sharp})	.
\end{align}
Combining (\ref{mwuselectionrule}) and (\ref{submwu}),
\begin{align}\label{interboundmwu}
\sum_{i=1}^{d}{w^{(t)}_{i}c_{i}(e^{(t+1)})}\leq &\frac{f_{T^{(t)}\cup S^{\sharp}}(e^{(t+1)})}{f(\mathrm{OPT}^{\prime} \cup S^{\sharp})-f(T^{(t)} \cup S^{\sharp})}\cdot \sum_{i=1}^{d}{w^{(t)}_{i}c_{i}(\mathrm{OPT}^{\prime})}\notag\\
\leq & \frac{f_{T^{(t)}\cup S^{\sharp}}(e^{(t+1)})}{f(\mathrm{OPT}^{\prime} \cup S^{\sharp})-f(T^{(t)} \cup S^{\sharp})}\cdot \sum_{i=1}^{d}{b^{\prime}_{i}w^{(t)}_{i}}.
\end{align}

Observe that 
\begin{align}
\sum_{i=1}^{d}{b^{\prime}_{i}w^{(t+1)}_{i}}=&\sum_{i=1}^{d}{b^{\prime}_{i}w^{(t)}_{i}\cdot \Big[1+\frac{\delta W c_{i}(e^{(t+1)})}{b^{\prime}_{i}}+\Big(\frac{\delta W c_{i}(e^{(t+1)})}{b^{\prime}_{i}}\Big)^{2}\Big]}\\
\leq & \sum_{i=1}^{d}{b^{\prime}_{i}w^{(t)}_{i}}+(\delta W+\delta^{2} W)\cdot \sum_{i=1}^{d}{w^{(t)}_{i}c_{i}(e^{(t+1)})}
\end{align}
where we use $Wc_{i}(e^{(t)})\leq b^{\prime}_{i}$ as $W=\min \{\frac{b^{\prime}_{i}}{c_{i}(e)}|e\in E\setminus (E_{\Gamma}\cup S^{\sharp}),i\in [d]\}$.	 Let $\Phi(t)=\sum_{i=1}^{d}{b^{\prime}_{i}w^{(t)}_{i}}$, then $\Phi(0)=d$, and 
\begin{align}\label{weightincrebound}
\Phi(t+1)\leq (1+\delta+\delta^{2})\cdot \Phi(t),
\end{align} 
which follows from the update rule of $w^{(t)}_{i}$ in Algorithm~\ref{linearbinary} and definition of $W$. Further we can obtain the following bound on the increment of $\Phi$ by utilizing (\ref{interboundmwu}),  
\begin{align}
\frac{\Phi(t+1)}{\Phi(t)}\leq & 1+ (\delta W+\delta^{2}W)\cdot \frac{f_{T^{(t)}\cup S^{\sharp}}(e^{(t+1)})}{f(\mathrm{OPT}^{\prime} \cup S^{\sharp})-f(T^{(t)} \cup S^{\sharp})}\notag\\
\leq & \exp\Big\{(\delta W+\delta^{2}W)\cdot \frac{f(T^{(t+1)} \cup S^{\sharp})-f(T^{(t)} \cup S^{\sharp})}{f(\mathrm{OPT}^{\prime} \cup S^{\sharp})-f(T^{(t)} \cup S^{\sharp})}\Big\},
\end{align}
which further implies that 
\begin{align}
\frac{\Phi(\bar{t})}{\Phi(0)}\leq &\exp\Big\{ (\delta W+\delta^{2}W)\cdot \sum_{t=0}^{\bar{t}-1} \frac{f(T^{(t+1)} \cup S^{\sharp})-f(T^{(t)} \cup S^{\sharp})}{f(\mathrm{OPT}^{\prime} \cup S^{\sharp})-f(T^{(t)} \cup S^{\sharp})} \Big\}\notag\\
\leq & \Big(\frac{f(\mathrm{OPT}^{\prime} \cup S^{\sharp})}{f(\mathrm{OPT}^{\prime} \cup S^{\sharp})-f(T^{(\bar{t})} \cup S^{\sharp})}\Big)^{(\delta W+\delta^{2}W)} 
\end{align}
Rearranging the terms, we have
\begin{align}
f(T^{(\bar{t})} \cup S^{\sharp})	\geq & \Big(1-\Big(\frac{\Phi(0)}{\Phi(\bar{t})}\Big)^{\frac{1}{\delta W+\delta^{2}W}} \Big) f(\mathrm{OPT}^{\prime} \cup S^{\sharp})\notag\\
\geq & \Big(1-\Big(\frac{d[1+\delta+\delta^{2}]}{e^{\delta W}}\Big)^{\frac{1}{\delta W+\delta^{2}W}}\Big)f(\mathrm{OPT}^{\prime} \cup S^{\sharp}),
\end{align}
For the second inequality, we utilize the fact that $\Phi(\bar{t}+1)\geq e^{\delta W}$ and $\Phi(\bar{t}+1)\leq \Phi(\bar{t})\cdot (1+\delta+\delta^{2})$. Note that
\begin{align*} 
\Big(\frac{d[1+\delta+\delta^{2}]}{e^{\delta W}}\Big)^{\frac{1}{\delta W+\delta^{2}W}}=&e^{\frac{-1}{1+\delta}}\cdot d^{\frac{1}{\delta W+\delta^{2}W}}\cdot (1+\delta+\delta^{2})^{\frac{1}{\delta W+\delta^{2}W}}\\
\leq & \frac{(1+2\delta)(1+\frac{2}{W})(1+\frac{2\log d}{\delta W})}{e} \leq \frac{1+15\delta}{e},
\end{align*}
where we use $e^{x}\leq 1+2x$ for $\forall x\in [0,1]$, and $W=\frac{2\log d}{\delta^{2}}$. The proof is complete.
\end{proofof}

\subsection{Proof}\label{appendixclaim1}
\begin{proof}
Utilizing the $\gamma$-approximate algorithm for $U^{\sharp}$-residual problem, we can obtain a set $U^{\prime}$ such that 
\begin{align}\label{residualapprox}
f_{U^{\sharp}}(U^{\prime})\geq \gamma \cdot \max_{S\subseteq E\setminus U^{\sharp}}f_{U^{\sharp}}(S) \geq \gamma f_{U^{\sharp}}(\mathrm{OPT}\setminus (U\cup U^{\sharp})),	
\end{align}
where the last inequality follows from the fact that $\mathrm{OPT}\setminus (U\cup U^{\sharp})$ is a feasible solution to $U^{\sharp}$-residual problem. Plugging the definition of $U^{\sharp}$-residual function into (\ref{residualapprox}), we can obtain
\begin{align*}
f(U^{\prime}\cup U^{\sharp})\overset{}{\geq} &\gamma f((\mathrm{OPT}\cup U^{\sharp})\setminus U)+(1-\gamma)f(U^{\sharp})\\
\overset{}{\geq} & \gamma f(\mathrm{OPT})+(1-\gamma\beta-\gamma)f(U^{\sharp}).
\end{align*}
If $\gamma+\gamma\beta\leq 1$, then we have $f(U^{\prime}\cup U^{\sharp})\geq \gamma f(\mathrm{OPT})$. Otherwise utilizing the simple fact that $f(U^{\sharp}) \leq f(U^{\prime}\cup U^{\sharp}) $, we can obtain $f(U^{\prime}\cup U^{\sharp})\geq \frac{f(\mathrm{OPT})}{1+\beta}$.
\end{proof}

\section{Supplementary Materials of Section~\ref{psystemdknapsection}}

\subsection{Algorithm for monotone objective}\label{appendixalgpsystemdknapsack}
\begin{algorithm}[H]\label{mainalgorithmpsystemdknapsack}
\small
    \caption{Main algorithm for monotone submodular function and $p$-system+$d$ knapsack constraints}
\textbf{Initialization:} $T \leftarrow \varnothing$, $\beta\leftarrow $ output of Algorithm \ref{btadt}, $\lambda\leftarrow 7(p+d+1)\cdot\beta$\\
\While{$\lambda \geq \frac{\beta}{p+7/4d+1}$}
{$T^{\prime}\leftarrow$ set returned by $\mathrm{BT}(\lambda,\frac{\delta}{p+1})$\\
$T\leftarrow T \cup \{T^{\prime}\}$\\
$\lambda \leftarrow \frac{\lambda}{1+\delta}$\\
}
\textbf{Return} $S_{o}\leftarrow\argmax_{S\in T}{f(S)}$
\end{algorithm}

\subsection{Details of Algorithm~\ref{btadt}}\label{appendixbtadt}
Similar to our treatments for cardinality constraint, we use the adaptive decreasing threshold algorithm to approximate the value of $f(\mathrm{OPT})$ in the following Algorithm~\ref{btadt}.

\begin{algorithm}[H]\label{btadt}
\small
    \caption{ADT for $\mathcal{I}=(\cap_{i=1}^{d}{\mathcal{K}_{i}})\cap \mathcal{I}_{p}$}
\textbf{Initialization:} $\underline{\omega}_1\leftarrow \frac{3\max_{e\in E}{f(e)}}{7(p+d+1)}$, $\bar{\omega}_1\leftarrow n\cdot \max_{e\in E}{f(e)}$, $\ell \leftarrow \lceil\log \log n\rceil$, $\epsilon=\frac{2}{7(p+1)(p+d+1)}$\\
\For{$i=1:\ell$}
{$\alpha_i=\exp(\log n\cdot e^{-i})-1$, $\lambda=\underline{\omega}_{i}$\\
\While{$\lambda \leq \bar{\omega}_{i}$}
{$S^{(i)}_{\lambda} \leftarrow \mathrm{BT}(\lambda,\epsilon)$, $\lambda \leftarrow \lambda(1+\alpha_{i})$\\
}
$\underline{\omega}_{i+1} \leftarrow  \frac{3\max_{\lambda} f(S^{(i)}_{\lambda})}{2}$, $\bar{\omega}_{i+1}\leftarrow \underline{\omega}_{i+1}(1+\alpha_{i})$\\
}
\textbf{Return} $\beta=\underline{\omega}_{\ell}$.
\end{algorithm}

\subsection{Proof of Proposition~\ref{psystemlemma}}\label{appendixpsystem}
\begin{proof}
The proof is the same as the analysis of greedy for maximizing a monotone submodular function under $p$-system constraint~\cite{badanidiyuru2014fast, calinescu2011maximizing}, here we provide the proof for completeness. An important note is that, for any element $e \in \mathrm{OPT}_{2}$, the reason that it cannot be added into $S$ is either the marginal increment of $e$ is less than $\frac{\varepsilon \max_{e}f(e)}{n}$, or $S+e$ is not a feasible set in $\mathcal{I}_{p}$. Owing to this observation, we are able to bound $R_{2}$ via same arguments for the analysis of the standard greedy for $p$-system constraint~\cite{calinescu2011maximizing,badanidiyuru2014fast}. Here we provide the proof for completeness.

For $S=\{e_{1}, e_{2},\ldots, e_{s}\}$ and $\forall i \in [s]$, we define set $C_{i}\subseteq \mathrm{OPT}_{2}\setminus S$ as,
\begin{align}
C_{i}=\Big\{e\in \mathrm{OPT}_{2}\setminus S \Big| S^{(i-1)}\cup \{e\}\in \mathcal{I} \Big\},	
\end{align}
which consists of the elements in $\mathrm{OPT}_{2}$ that are able to be added into the candidate solution set in the $i$-th step. According to the down-closed property of the independent system, we know that $C_{i+1}\subseteq C_{i}$ and we have $C_{1}=\mathrm{OPT}_{2}\setminus S$. Consider set $Q_{i}=S^{(i)}\cup (C_{1}\setminus C_{i+1})$. On the one hand, we have $C_{1}\setminus C_{i+1}\in \mathcal{I}$, since it is a subset of $\mathrm{OPT}\in \mathcal{I}$, which implies that $Q_{i}$ has a base of size no less than $|C_{1}\setminus C_{i+1}|$. On the other hand, $S^{(i)}$ is a base of $Q_{i}$ since no elements in $Q_{i}\setminus S^{(i)}$ can be added into $S^{(i)}$ according to the definition of $C_{i}$. Then based on the definition of $p$-system, we know that
\begin{align}\label{sizeofA}
|C_{1}\setminus C_{i+1}|\leq p \cdot |S^{(i)}|=pi.	
\end{align}
Now consider the procedure of decreasing threshold $\Delta$. For $1\leq i \leq s$, we let $\Delta_{i}$ be the value of $\Delta$ when element $e_{i}$ is added into $S$, then we have 
\begin{align}
f(S^{(i)})-f(S^{(i-1)})\geq \Delta_{i}.	
\end{align}
And we further claim that
\begin{align}\label{thresholdpsystem}
f(S^{(i-1)}+e)-f(S^{(i-1)})\leq \frac{\Delta_{i}}{1-\varepsilon}=\Delta_{i-1},\;\forall e\in C_{i}.	
\end{align}
Otherwise $e \in C_{i}$ satisfying (\ref{thresholdpsystem}) will already be included into the candidate solution set in previous iteration since
\begin{itemize}
\item $S^{(i-1)}+e\in \mathcal{I}$ is a feasible set according to the definition of $C_{i}$;
\item $f(S^{(i-1)}+e)-f(S^{(i-1)})\geq \theta \cdot [\sum_{i=1}^{d}c_{i}(e)]$ holds, which is based on the submodularity and the definition of $\mathrm{OPT}_{2}$;
\item The marginal increment of $e$ with respect to set $S^{(i-1)}$ is no less than the threshold $\Delta_{i-1}$ according to (\ref{thresholdpsystem}).
\end{itemize}
However $e\notin S$, thus (\ref{thresholdpsystem}) is true. Using similar arguments, we obtain
\begin{align}\label{minthresholdpsystem}
f(S+e)-f(S)\leq \frac{\varepsilon}{n}M,\;\forall e\in C_{s+1}.
\end{align}
Hence we are able to show that 
\begin{align}
R_{2}=&f(S\cup \mathrm{OPT}_{2})-f(S)\\
\overset{}{\leq} & \sum_{e\in \mathrm{OPT}_{2}\setminus S}{\Big[f(S+e)-f(S)\Big]}\tag{submodularity}\\
\overset{(a)}{=} & \sum_{i=1}^{s}{\sum_{e\in C_{i}\setminus C_{i+1}}{\Big[f(S+e)-f(S)\Big]}}+\sum_{e\in C_{s+1}}{\Big[f(S+e)-f(S)\Big]}  \\ 
\overset{(b)}{\leq} & \frac{1}{1-\varepsilon}\sum_{i=1}^{s}{|C_{i}\setminus C_{i+1}|\cdot\Delta_{i}}+|C_{r+1}|\cdot \frac{\varepsilon}{n}M,\label{lastinesystem}
\end{align}
where $(a)$ is based on the definition of $A_{i}$ and $(b)$ follows from inequalities (\ref{thresholdpsystem})-(\ref{minthresholdpsystem}). Observe that 
\begin{itemize}
\item $\{\Delta_{i}\}_{i\in [s]}$ is a decreasing sequence;
\item The total sum of sequence $\{|C_{i}\setminus C_{i+1}|\}_{i\in [s]}$ are fixed;
\item $|C_{i}\setminus C_{i+1}|\leq p$, according to (\ref{sizeofA}).
\end{itemize}
Hence $\sum_{i=1}^{s}{|C_{i}\setminus C_{i+1}|\cdot \Delta_{i}}$ achieves its maximum when $|C_{i}\setminus C_{i+1}|=p$. As a consequence, the following upper bound holds for the first term in (\ref{lastinesystem}),
\begin{align}\label{upboundfirstterm}
\sum_{i=1}^{s}{|C_{i}\setminus C_{i+1}|\cdot \Delta_{i}}\leq \sum_{i=1}^{s} {p\cdot\Delta_{i}}\leq p\cdot f(S).	
\end{align}
Now plugging (\ref{upboundfirstterm}) and inequality $|C_{r+1}|\cdot \frac{\varepsilon}{n}M\leq M\leq f(\mathrm{OPT})$ into (\ref{lastinesystem}), the proof is complete.	
\end{proof}

\subsection{Proposition~\ref{wlboundpd} and proof}\label{appendixwlboundpd}

Algorithm~\ref{btadt} returns $\underline{w}_{\ell}$, which is shown to be a good approximation of $f(\mathrm{OPT})$ in Proposition~\ref{wlboundpd}. 
\begin{proposition}\label{wlboundpd} For any $i\geq 0$, we have
\begin{align}\label{psystembound}
\frac{3 f(\mathrm{OPT})}{7(p+d+1)(1+\alpha_{i})}\leq \underline{w}_{i+1}\leq \frac{3}{2}f(\mathrm{OPT}),
\end{align}
which implies that $\underline{w}_{\ell}\in [\frac{f(\mathrm{OPT})}{7(p+d+1)(1+c)}, \frac{3f(\mathrm{OPT})}{2}]$.
\end{proposition}

\begin{proof}
We first note that 
\begin{align}\label{pdsytemineq}
f(S^{*})\geq &\min \Big \{ \frac{2\theta}{3}, \frac{f(\mathrm{OPT})-\theta (d-|\mathrm{OPT}\cap B|/2)}{p+|\mathrm{OPT}\cap B|+1}-(p\varepsilon+\varepsilon)\cdot f(\mathrm{OPT})\Big \}.
\end{align}
Inequality~(\ref{pdsytemineq}) directly follows from Proposition \ref{tildeeexist} and \ref{propositioneexists}, by taking the minimum of the two lower bounds. The RHS of (\ref{psystembound}) directly follows from the fact that 
\begin{align*}
\underline{w}_{i+1}=\frac{3}{2}\max_{\lambda}f(S^{(i)}_{\lambda})\leq \frac{3}{2}f(\mathrm{OPT}).    
\end{align*}
We finish the proof of the LHS by induction. For the base case when $i=0$, inequality (\ref{psystembound}) is equivalent to the statement that $\underline{w}_{1}\geq \frac{3f(\mathrm{OPT})}{7(p+d+1)n}$, which is true based on the definition of $\underline{w}_{1}$. Now suppose that (\ref{psystembound}) holds for $i=s-1$, \ie,
\begin{align}\label{inductionassuppd}
\underline{w}_{s}\geq \frac{3f(\mathrm{OPT})}{7(p+d+1)(1+\alpha_{s-1})}. 
\end{align} 
Following from (\ref{inductionassuppd}), we have $\bar{w}_{s}=(1+\alpha_{s-1})\underline{w}_{s}\geq \frac{3f(\mathrm{OPT})}{7(p+d+1)}$. As a consequence, there must exist an integer $z_{s}$ during the $s$-th iteration, such that 
\begin{align}\label{boundoflambda}
\lambda^{*}_{s}=\underline{\omega}_{s}(1+\alpha_{s})^{z_{s}}\in \Big[\frac{3f(\mathrm{OPT})}{7(p+d+1)(1+\alpha_{s})}, \frac{3f(\mathrm{OPT})}{7(p+d+1)}\Big],
\end{align}
which implies that $\underline{w}_{s+1}$ can be lower bounded as follows when $\varepsilon=\frac{2}{7(p+1)(p+d+1)}$,
\begin{align}
\underline{w}_{s+1}\overset{(a)}{\geq} & \frac{3f(BT(\lambda^{*}_{s}))}{2}\notag\\
\overset{(b)}{\geq} & \frac{3}{2}\cdot \min\Big \{ \frac{2}{3}\cdot \frac{3f(\mathrm{OPT})}{7(1+\alpha_{s})(p+d+1)}, \frac{f(\mathrm{OPT})-\frac{3(d-|\mathrm{OPT}\cap B|/2)OPT}{7(p+d+1)}}{p+|\mathrm{OPT}\cap B|+1}-\frac{2f(\mathrm{OPT})}{7(p+d+1)}\Big \}	\notag\\
\overset{(c)}{\geq} & \frac{3}{7(p+d+1)(1+\alpha_{s})} f(\mathrm{OPT}) \notag
\end{align}
where $(a)$ is based on the definition of $\underline{w}_{s+1}$. Plugging (\ref{boundoflambda}) into (\ref{pdsytemineq}), we can obtain $(b)$. $(c)$ follows from Fact~\ref{numberofbigelebound}. Hence we have 
\begin{align*}
\bar{w}_{s+1}=(1+\alpha_{s})\cdot \underline{w}_{s+1}\geq \frac{3}{7(p+d+1)} f(\mathrm{OPT}),
\end{align*}
which indicates that (\ref{psystembound}) also holds for $i=s$. The proof is complete.		
\end{proof}

\subsection{Proof of Theorem~\ref{psystemtheorem}}\label{appendixlemmapsystem}
\begin{proof}
Let the optimal threshold
\begin{align*}
\theta^{*}=\frac{f(\mathrm{OPT})}{d+2/3+|\mathrm{OPT}\cap B|/6+\frac{2}{3}p}.
\end{align*}
According to Proposition \ref{wlboundpd}, it is easy to see that 
\begin{align*}
\theta^{*}\in \Big[\frac{3}{2}\cdot\frac{f(\mathrm{OPT})}{p+\frac{7}{4}d+1}, f(\mathrm{OPT})\Big] \subseteq \Big[\frac{w_{\ell}}{p+7/4d+1} ,7(p+d+1)(1+c)\underline{w}_{\ell}\Big]. 
\end{align*}
Hence there exist an iteration in which $\lambda \in [(1-\delta)\theta^{*},\theta^{*}]$, from which we know that
\begin{align*}
f(S_{o})\geq & \min \Big \{ \frac{2(1-\delta)\theta^{*}}{3}, \frac{f(\mathrm{OPT})-\theta^{*}(d-|\mathrm{OPT}\cap B|/2)}{p+|\mathrm{OPT}\cap B|+1}-\delta f(\mathrm{OPT})\Big \}\\
\geq & \Big(\frac{1}{p+7d/4+1}-2\delta \Big)\cdot f(\mathrm{OPT}).
\end{align*}
Note that using our adaptive decreasing threshold algorithm, we are able to obtain a constant approximation of $f(\mathrm{OPT})$ in $\log\log n$ rounds, while the time complexity in each round is $n\log n$, thus the total time complexity is $O_{\varepsilon}(n\log n \cdot\log\log n)$. The proof is complete.	
\end{proof}


\subsection{Details of the Algorithm for non-monotone objective and $p$-system+$d$ knapsack}\label{algononmonotonepd}

\begin{algorithm}[H]
\label{nonmonotoneadt}
\small
    \caption{Maximizing non-monotone submodular function under $p$-system+$d$ knapsack constraints}
\textbf{Input:} Algorithm BT($\theta,\varepsilon$).\\
\textbf{Output:} A constant approximation of $f\mathrm{OPT})$\\
\textbf{Initialization:} $\underline{\omega}_1\leftarrow \frac{3\max_{e\in E}{f(e)}}{(8p+7d+4)}$, $\bar{\omega}_1\leftarrow n\cdot \max_{e\in E}{f(e)}$, $U\leftarrow \varnothing$, $\ell \leftarrow \log\log n $.\\
\For{$i=1:\ell$}
{$\alpha_i=\exp(\log n\cdot e^{-i})-1$\\
$\lambda=\underline{\omega}_{i}$\\
\While{$\lambda \leq \bar{\omega}_{i}$}
{$S^{(i)}_{\lambda} \leftarrow \mathrm{BT}(\lambda,\frac{2}{(p+1)(8p+7d+4)})$\\
$\lambda \leftarrow \lambda(1+\alpha_{i})$\\
}
$\underline{\omega}_{i+1} \leftarrow  \frac{3\max_{\lambda} f(S^{(i)}_{\lambda})}{2}$\\
$\bar{\omega}_{i+1}\leftarrow \underline{\omega}_{i+1}(1+\alpha_{i})$\\
}
\textbf{Return} $\omega_{\ell}$.
\end{algorithm}

\end{document}